\documentclass[reprint,amsmath,twocolumn,amssymb, nobibnotes, aps, pra, superscriptaddress]{revtex4-1}

\usepackage{graphicx}% Include figure files
\usepackage{dcolumn}% Align table columns on decimal point
\usepackage{bm}% bold math
\usepackage{color}
\usepackage{tikz}
\usetikzlibrary{shapes}
\usetikzlibrary{backgrounds}
\usetikzlibrary{plotmarks}
\usepackage{MnSymbol,wasysym}
\usepackage{gensymb}
\usepackage{mathtools}
\usepackage{comment}

\newcommand{\RR}{{\mathbb R}}

\begin{document}
\def \beq{\begin{equation}}
\def \eeq{\end{equation}}
\def \bea{\begin{eqnarray}}
\def \eea{\end{eqnarray}}
\def \bem{\begin{displaymath}}
\def \eem{\end{displaymath}}
\def \P{\Psi}
\def \Pd{|\Psi(\boldsymbol{r})|}
\def \Pds{|\Psi^{\ast}(\boldsymbol{r})|}
\def \Po{\overline{\Psi}}
\def \bs{\boldsymbol}
\def \bl{\bar{\boldsymbol{l}}}
\newcommand{\ihat}{\hat{\textbf{\i}}}
\newcommand{\jhat}{\hat{\textbf{\j}}}

\title{Chiral solitary waves in  a nonlinear topological insulator model}

\author{Troy I. Johnson}
\email{tjohnso3@uccs.edu}
\affiliation{Department of Mathematics, University of Colorado, Colorado Springs, Colorado 80918, USA}

\author{Justin T. Cole}
\email{jcole13@uccs.edu}
\affiliation{Department of Mathematics, University of Colorado, Colorado Springs, Colorado 80918, USA}

\begin{abstract}

 An outstanding challenge in the field of topological insulators is the realization of  nonlinear systems that support coherent traveling  waves. 
 Highly nonlinear lattices can suffer from significant radiation losses due to Peierls-Nabarro effects. In this work a nonlinear tight-binding model that supports robust traveling edge states is proposed and examined. This system possess a nontrivial local Chern topology and  soliton-like states. When a traveling solitary wave collides with a stationary mode, the two are observed to interact inelastically.  These results  suggest future directions for the modeling, realization, and application
 of nonlinear Chern insulators.

\end{abstract}
\maketitle

%%%%%%%%%%%%%%%%%%%%%%%%%%%%%%%%%%%%%%%%%%%%%%%%%%%%%%%%

\section{Introduction}

Topological insulators are states which are not entirely conductors nor insulators. Along the boundary of a topological medium they conduct, but in their interior they act as insulators. These systems are characterized by topological invariants that are preserved under continuous perturbations of the system. As such, they are expected to be resilient to perturbations.
This connection between bulk topological invariants and presence of edge states is typically referred to as the bulk-edge correspondence.

Topological insulators have been observed in a wide variety of physical systems. Originally discovered in condensed matter physics \cite{vonKlitzing}, they have since been realized  
in acoustics \cite{CNRSgroup}, photonics \cite{Rechtsman,Ozawa}, gryomagnetic crystals \cite{Wang}, and ultracold fermionic systems \cite{Jotzu2014}, to name a few. The diversity of physical contexts speaks to the universality of these systems and their importance. %More comprehensive reviews can be found in \cite{Ozawa}.

Chern insulators are topological insulators associated with a nontrivial topological invariant,   the Chern number \cite{Thouless}. The edge states are chiral in nature and in 2D propagate clockwise or counterclockwise. %, depending on the value of the Chern number. 
The quintessential minimal model for Chern insulators is the Haldane model \cite{Haldane1988},  an %and is an
 effective reduced-order model in numerous physical systems \cite{Jotzu2014,Nixon2023,Ablowitz24,Lannebere}. The model is simple enough to solve, yet rich enough to display nontrivial topological features.
%This  discrete model captures the essence of breaking time-reversal symmetry for the realization of nonreciprocal propagation.

To date, most research has focused on linear topological insulators. However, it is natural to examine the effect nonlinearity may have on these systems when high power or amplitudes become non-negligible. Nonlinear effects such as high-harmonic generation \cite{Boyd} and solitons \cite{KivsharBook} have previously been studied in other photonic systems. Can one realize nonlinear coherent structures with nontrivial topology  in Chern insulators? In the case of bulk solitons, the answer is yes, as they have been theoretically and experimentally observed in Floquet systems \cite{Rechtsman2013,Rechtsman2020}. In those works cyclonic steady-state solutions were found to exist. Modulational instability has also been reported \cite{Kartashov2016,Leykam2021}.

What is also desirable are systems that support genuine traveling edge modes, like their linear counterparts. Several nonlinear models have been considered. In \cite{Leykam,Cerj2023}, a modification of the Haldane mass term to include a saturable-type nonlinearity was studied. In doing so, the system looks linear in the small amplitude limit and massless in the high intensity limit. Another notable work \cite{Isobe24} has  established a nonlinear version of the bulk-edge correspondence in systems with dispersive permittivity.  There, the  nonlinearity considered is %was 
an extended gyrotropic permittivity and not the %cubic 
intensity dependent type we consider here. Another work has demonstrated a  bulk-edge correspondence in a nonlinear Qi-Wu-Zhang  model \cite{Sone24} using a rescaled Chern number. Several thorough review articles give more examples and insight \cite{LeykamReview, AblowitzReview,KangReview,RechtsmanReview}

% Note that in spite of this having a provable nontrivial topological invariant, soliton wave packets still do not form.

For the case of on-site Kerr type nonlinearities, the results have been mixed. Early results suggested edge solitons for slowly varying profiles in    photonic crystals \cite{Ablowitz2014,Chong2016,Kartashov2016}. However, in more practical  crystals these are challenging. %not feasible. 
Subsequent research has shown very nonlinear, highly localized initial states emit significant amounts of radiation \cite{Ablowitz2021,Rechtsman21}. These highly discrete systems experience  Peierls-Nabarro (PN) related effects. The PN barrier manifests  due  to the lack of  translational invariance in the model \cite{Kevrekidis2009}. As such, solitary waves have different local energy, depending on whether their peaks are on-site vs. off-site, for example \cite{Kivshar,Jenkinson2015}. In a nonlinear Floquet insulator \cite{Ablowitz2021} the mode can radiate  
until a  slowly-varying profile balances a typically weak nonlinearity.  

On the other hand, there are other discretizations that are known to support traveling solitons. Notably, the Ablowitz-Ladik equation is an integrable (exactly solvable) model with soliton solutions \cite{Ablowitz1975,Ablowitz1976}. The model is characterized by slightly nonlocal nonlinearities which in the long-wave limit reduce to the 1D nonlinear Schr\"odinger  equation. The Ablowitz-Ladik model has been considered in a wide variety of physical applications, including: two-level atomic systems \cite{Konotop1997}, waveguide arrays \cite{Aceves1996}, spin Heisenberg models \cite{Haldane1982,Ishimori1982} and topoelectric circuits \cite{Castro2025}.
%The model is equivalent to certain  %and can be obtained by a reduction of orthogonal polynomials on the unit circle \cite{Nenciu2005}. 
This is the inspiration for this work.

In this article we introduce a Haldane model with nonlinearity  that is inspired by the Ablowitz-Ladik equation.  The model contains two conserved quantities and is considerably more effective at supporting traveling edge states than the standard on-site nonlinearity. By colliding %the 
traveling states and stationary solitons, we observe an  inelastic interaction. The significance of this work is that it suggests future directions for the realization of %the
nonlinear Chern insulator systems and edge solitons.

This is not the first work to consider Ablowitz-Ladik type nonlinearity in the context of topological insulators. One-dimensional discrete models  were considered in \cite{Munoz2017} and \cite{Castro2025}. The former explored nonlinear switching in a two sublattice system with twisted boundary conditions, while the latter examined nonreciprocal motion induced by an intrinsic instability. Our work appears to be the first to consider the Ablowitz-Ladik nonlinearity in a 2D topological insulator system with nonreciprocal motion. 

The outline of the work is as follows. In Section \ref{Haldane_sec} the linear Haldane model, its topology and localized edge solutions are reviewed. In Section \ref{NL_Haldane_Sec} two nonlinear models that build off the Haldane model are introduced.  Conserved  quantities and a topology characterization are given. In Section \ref{Travel_Sec} the traveling edge current dynamics in the two models are compared. A family of stationary surface solitons is also found. In Section \ref{interact_sec} the interaction of traveling solitonlike edge modes and stationary solitons is explored. We conclude in Section \ref{conclude_sec}.

%%%%%%%%%%%%%%%%%%%%%%%%%%%%%%%%%%%%%%%%%%%%%%%%%%%%%%%%%%%
\section{The Haldane model}
\label{Haldane_sec}
%%%%%%%%%%%%%%%%%%%%%%%%%%%%%%%%%%%%%%%%%%%%%%%%%%%%%%%%%%%

%%%%%%%%%%%%%%%%%%%%%%%%%%%%%%%%%%%%%%%%
\begin{figure}
\centering
  \includegraphics[scale=0.39]{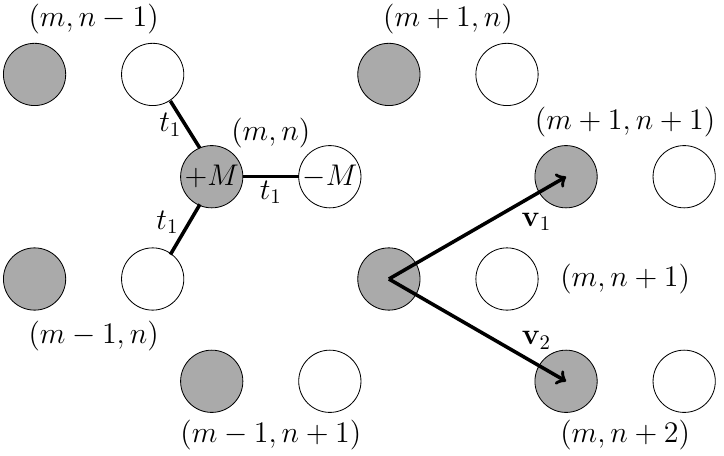} \hspace{0.1 in}
    \includegraphics[scale=0.4]{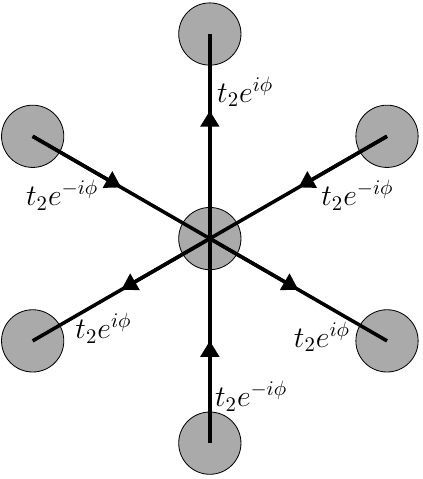}
\caption{The honeycomb lattice with a-sites (gray discs) and b-sites (white discs). The mass term $\pm M \in \RR$ denotes an inversion breaking self-interaction for the a-sites and b-sites, respectively, $t_1 \in \mathbb{R}$ denotes nearest neighbor coupling,  $t_2 \ge 0 $ is the magnitude of the next-nearest neighbor coupling and $\phi \in (-\pi,\pi]$ is the local magnetic potential, time-reversal parameter. \label{HC_lattice_diagram}}
 \end{figure}
%%%%%%%%%%%%%%%%%%%%%%%%%%%%%%%%%%%%%%%%

  Consider a honeycomb lattice with lattice vectors 
   ${\bf v}_1 = ( 3/2,  \sqrt{3}/2 ),  {\bf v}_2= (3/2,  -\sqrt{3}/2 ),   $ scaled so that the lattice vector lengths are $||{\bf v}_1||=||{\bf v}_2||=\sqrt{3}$. %$|{\bf v}_1|=|{\bf v}_2|=\sqrt{3}$. 
   The   $a$-sites are located at $m {\bf v}_1+n {\bf v}_2 $ and $b$-sites %located
   at $m {\bf v}_1+n {\bf v}_2+ {\bf d}$ for $m,n \in \mathbb{Z}$ and
   ${\bf d}=(1, 0)$ so that the nearest neighbor distance is scaled to unity. The lattice vectors relative to the honeycomb lattice are shown in Fig.~\ref{HC_lattice_diagram}.

  Define the linear Haldane operators \cite{Haldane1988}, for the $a$-site and $b$-site equations, respectively, as
%%%%%%%%%%%%%%%%%%%%%%%%%%%%%%%%%%%%%%%%%%%%%%%%%%%%%%%%%%%%%%%%%%%%%%
\small
\begin{equation}
       \label{ha_sites}
       \begin{split}
       H_a[a_{mn},b_{mn}]\equiv M a_{mn}+ t_1\left(b_{mn}+b_{m-1,n}+b_{m,n-1}\right)\\+t_2e^{i\phi}\left(a_{m-1,n}+a_{m,n+1}+a_{m+1,n-1}\right)\\+t_2e^{-i\phi}\left(a_{m+1,n}+a_{m,n-1}+a_{m-1,n+1}\right) , 
   \end{split}
   \end{equation}
   \begin{equation}
       \label{hb_sites}
       \begin{split}
       H_b[a_{mn},b_{mn}] \equiv - M b_{mn}+ t_1\left(a_{mn}+a_{m+1,n}+a_{m,n+1}\right)\\+t_2e^{i\phi}\left(b_{m+1,n}+b_{m,n-1}+b_{m-1,n+1}\right)\\+t_2e^{-i\phi}\left(b_{m-1,n}+b_{m,n+1}+b_{m+1,n-1}\right) .
   \end{split}
   \end{equation}
%%%%%%%%%%%%%%%%%%%%%%%%%%%%%%%%%%%%%%%%%%%%%%%%%%%%%%%%%%%%%%%%%%%%%%
\normalsize
In a typical fashion, $\pm M \in \RR$ is a mass term which breaks inversion symmetry in the lattice, %self interaction for the a-sites (b-sites),
$t_1 \in \RR $ is the nearest neighbor coupling coefficient (from the 3 nearest sites), %gives the nearest neighbor interaction , 
and $t_2e^{\pm i \phi}$, $t_2 \ge 0, \phi \in (-\pi , \pi]$ denotes the complex next-nearest neighbor coefficient. % interaction in the form of a clockwise circulation. 
Physically, $\phi$ is related to the local magnetic vector potential %field strength 
such that the net magnetic flux over the unit cell %hexagon 
is zero. This latter term is responsible for breaking the time reversal symmetry of the model. As a result, the time-dependent linear Haldane system can be written succinctly as
%%%%%%%%%%%%%%%%%%%%%%%%%%%%%%%%%%%%%%%%%%%%%%%%%%%%%%%%%%%%%%%%%%%%%%
   \begin{equation}
       \label{ab_sites}
       i\frac{da_{mn}}{dt}=H_a[a_{mn},b_{mn}] , ~~~~~ i\frac{db_{mn}}{dt} = H_b[a_{mn},b_{mn}] .
   \end{equation}
%%%%%%%%%%%%%%%%%%%%%%%%%%%%%%%%%%%%%%%%%%%%%%%%%%%%%%%%%%%%%%%%%%%%%%

On an infinite domain with localized boundary conditions, look for  Fourier (bulk) solutions of the form
%%%%%%%%%%%%%%%%%%%%%%%%%%%%%%%%%%%%%%%%%%%%%%%%%%%%
%\begin{equation}
%\label{ab_fourier}
$        a_{mn}(t)=\alpha({\bf k})e^{i {\bf k} \cdot(m {\bf v}_1+n {\bf v}_2)-i\lambda t} ,     b_{mn}(t)=\beta({\bf k})e^{i {\bf k}\cdot(m {\bf v}_1+n {\bf v}_2+ {\bf d})-i\lambda t} .$
%\end{equation}
%%%%%%%%%%%%%%%%%%%%%%%%%%%%%%%%%%%%%%%%%%%%%%%%%%%%
Doing so, one obtains a $2 \times 2$ linear eigenvalue system $   \mathcal{L} {\bf u} = \lambda {\bf u}$ with ${\bf u}(k) = (\alpha(k), \beta(k))^T$.
%%%%%%%%%%%%%%%%%%%%%%%%%%%%%%%%%%%%%%%%%%%%%%%%%%%%%
%    \begin{equation}
%        \label{alpha_sites}
%        \lambda\alpha= t_1\sum_{j=1}^3e^{-ik\cdot a_j}\beta+t_2\left[e^{i\phi}\sum_{j=1}^3e^{-ik\cdot b_j}+e^{-i\phi}\sum_{j=1}^3e^{ik\cdot b_j}\right]\alpha
%    \end{equation}
%%%%%%%%%%%%%%%%%%%%%%%%%%%%%%%%%%%%%%%%%%%%%%%%%%%%%
%    \begin{equation}
%        \label{beta_sites}
%       \lambda\beta= t_1\sum_{j=1}^3e^{ik\cdot a_j}\alpha+t_2\left[e^{i\phi}\sum_{j=1}^3e^{ik\cdot b_j}+e^{-i\phi}\sum_{j=1}^3e^{-ik\cdot b_j}\right]\beta .
%    \end{equation}
%%%%%%%%%%%%%%%%%%%%%%%%%%%%%%%%%%%%%%%%%%%%%%%%%%%%%
%JTC: To parameterize $k$, we express it in terms of  the reciprocal lattice vectors \jtc{$k_1= 2\pi ( 1/3 , 1 / \sqrt{3} )$ and  $k_2= 2\pi ( 1/3, -1/ \sqrt{3})$}.}  
The eigenfunctions of this system can exhibit  a nonzero Chern number %, a topological invariant 
\cite{Haldane1988,Asboth2016}. Indeed, for any $t_1 \not=0, t_2> 0 $ and $\phi \not= n \pi$ (complex next-nearest neighbors coefficients), the bulk eigenvectors contain a nonzero Chern number when $|M| < 3 \sqrt{3} t_2  | \sin\phi |$. Applying the bulk-edge correspondence, one expects to find chiral edges states that propagate unidirecionally around the boundary of a finite domain. 
    
%%%%%%%%%%%%%%%%%%%%%%%%%%%%%%%%%%%%%%%%%%%%%%%%%%%%%%%%%%%%%%%%
%%%%%%%%%%%%%%%%%%%%%%%%%%%%%%%%%%%%%%%%%%%%%%%%%%%%%%%%%%%%%%%%
 \subsection{Linear Edge Modes}
 %%%%%%%%%%%%%%%%%%%%%%%%%%%%%%%%%%%%%%%%%%%%%%%%%%%%%%%%%%%%%%%%
 \label{linEdge}

 Edge mode solutions can be found by %effectively 
 taking a Fourier transform %periodic %or localized boundary conditions 
 in one direction and open (zero) boundary conditions in the other.  Look for time-harmonic, separable solutions of the form
%%%%%%%%%%%%%%%%%%%%%%%%%%%%%%%%%%%%%%%%%%%%%%%%%%%%%%%%%%%%%%%
 \begin{equation}
     \label{oneD_a_fourier}
     \begin{split}
     a_{mn}(t)=a_n ({\bf k})e^{i ( {\bf k} \cdot m {\bf v}_1 -  \lambda ( {\bf k}) t) } ,\\  b_{mn}(t) =b_n ({\bf k}) e^{i ( {\bf k} \cdot m {\bf v}_1 -  \lambda ( {\bf k}) t) } ,
     \end{split}
 \end{equation}
%%%%%%%%%%%%%%%%%%%%%%%%%%%%%%%%%%%%%%%%%%%%%%%%%%%%%%%%%%%%%%%
 where a Fourier transform has taken place in the ${\bf v}_{1}$ direction. As a result, the system in (\ref{ab_sites}) reduces to
%%%%%%%%%%%%%%%%%%%%%%%%%%%%%%%%%%%%%%%%%%%%%%%%%%%%%%%%%%%%%%%
\begin{equation}
       \label{eig_ab}
       \begin{split}
           \lambda({\bf k}) a_{n}({\bf k})= e^{- i {\bf k} \cdot m {\bf v}_1  } H_a[a_{n} e^{i {\bf k}\cdot m {\bf v}_1 },b_{n} e^{i {\bf k} \cdot m {\bf v}_1 }] ,\\\lambda({\bf k}) b_{n}({\bf k})= e^{-i {\bf k} \cdot m {\bf v}_1 } H_b[a_{n} e^{i {\bf k} \cdot m {\bf v}_1 } ,b_{n} e^{i {\bf k} \cdot m {\bf v}_1 }],
\end{split}
   \end{equation}
%%%%%%%%%%%%%%%%%%%%%%%%%%%%%%%%%%%%%%%%%%%%%%%%%%%%%%%%%%%%%%
%where 
such that all explicit dependence on $m$ is removed from the system. For some positive integer $N$, we impose open boundary conditions: $a_{n}=b_{n}=0$ for $n \leq 0$ or $n \geq N+1$. 
Thus we  solve (\ref{eig_ab}) on  a %semi-infinite 
lattice stripe which is large but finite in the ${\bf v}_2$ direction and infinite in the $ {\bf v}_1$ direction.

%%%%%%%%%%%%%%%%%%%%%%%%%%%%%%%%%%%%%%%%%%%%%%%%%%%%%%%%%%
  We can express ${\bf k}$ as a linear combination of the reciprocal lattice vectors ${\bf k}_1 $ and ${\bf k}_2 $. That is, ${\bf k}=r {\bf k}_1+s {\bf k}_2$ for $r,s\in \RR$. Observe that ${\bf k}_i \cdot {\bf v}_j=2\pi\delta_{ij}$, and so we have  ${\bf k} \cdot {\bf v}_1=(r {\bf k}_1+s {\bf k}_2)\cdot {\bf v}_1= %r {\bf k}_1\cdot {\bf v}_1 +s {\bf k}_2 \cdot {\bf v}_2=
  2\pi r$. We can simplify (\ref{eig_ab}) to
%%%%%%%%%%%%%%%%%%%%%%%%%%%%%%%%%%%%%%%%%%%%%%%%%%%%%%%%%%%%%%
\small
\begin{equation}
       \label{r_a}
       \begin{split}
\lambda(r) a_{n}= Ma_n+t_1\left(b_{n}+b_{n}e^{-2\pi r i}+b_{n-1}\right)\\+t_2e^{i\phi}\left(a_{n}e^{-2\pi ri}+a_{n+1}+a_{n-1}e^{2\pi r i}\right)\\+t_2e^{-i\phi}\left(a_{n}e^{2\pi r i }+a_{n-1}+a_{n+1}e^{-2\pi r i }\right) ,
   \end{split}
   \end{equation}
%%%%%%%%%%%%%%%%%%%%%%%%%%%%%%%%%%%%%%%%%%%%%%%%%%%%%%%%%%%%
   \begin{equation}
       \label{r_b}
       \begin{split}
       \lambda(r) b_{n}=-Mb_n+t_1\left(a_{n}+a_{n}e^{2\pi ri }+a_{n+1}\right)\\+t_2e^{-i\phi}\left(b_{n}e^{2\pi r i }+b_{n-1}+b_{n+1}e^{-2\pi r i}\right)\\+t_2e^{i\phi}\left(b_{n}e^{-2\pi r i }+b_{n+1}+b_{n-1}e^{2\pi r i }\right) , 
   \end{split}
   \end{equation}
%%%%%%%%%%%%%%%%%%%%%%%%%%%%%%%%%%%%%%%%%%%%%%%%%%%%%%%%%%%%%%
\normalsize
with %where 
$r \in [0, 1)$ due to periodicity.

%%%%%%%%%%%%%%%%%%%%%%%%%%%%%%%%%%%%%%%%%%%%%%%%%%%%%%%%%%%%%%
      \begin{figure*}
   \centering
\includegraphics[scale=.35]{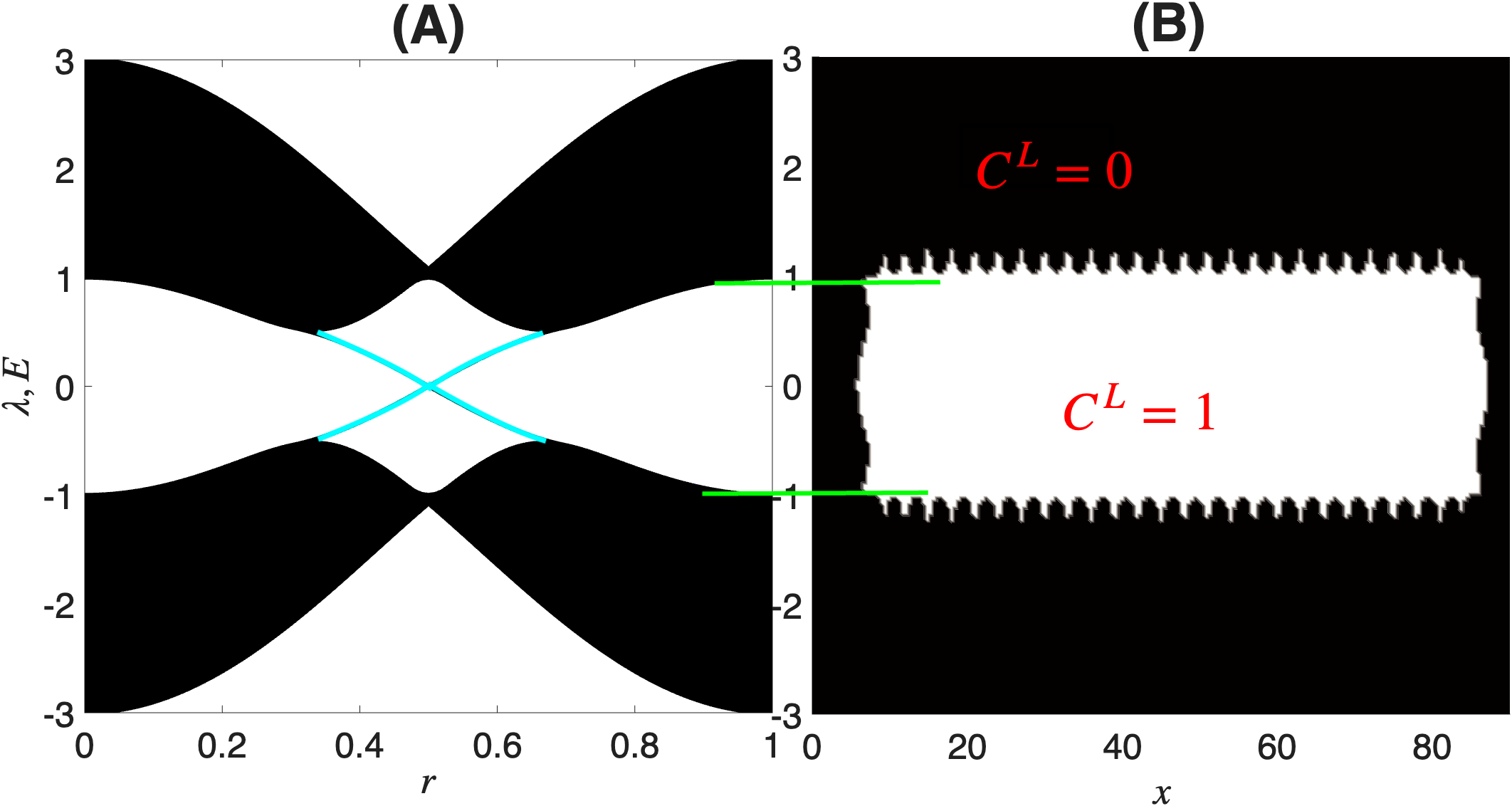}
\caption{(A) Spectral bands obtained by solving  edge problem (\ref{r_a})-(\ref{r_b}) on a semi-infinite interval with N = 121 sites. The middle gap (between the green lines) lies between $-1 \le  \lambda \le  1$; cyan lines are $\lambda$ values that correspond to edge modes.
The parameters used are  $M=0$, $t_1=1$, $t_2=0.1$, $\phi=\pi/2$. (B) Local Chern number (\ref{local_chern}) of the Haldane system, (\ref{ab_sites}), at $y=0$ for  various $x$ and $E$, %$\lambda$, 
with $N=30$, $\kappa_y=0.1$ and $\kappa_x=\kappa_y/\sqrt{3}$.  White (black) corresponds to a local Chern number of $1 (0)$. A spatial scan of the local topology is shown in Fig.~\ref{haldaneY}.\label{linearbands}}
   \end{figure*}
%%%%%%%%%%%%%%%%%%%%%%%%%%%%%%%%%%%%%%%%%%%%%%%%%%%%%%%%%%%%%%

   Now let ${\bf c}=(a_1,a_2,...,a_N|b_1,b_2,...,b_N)^T$ denote our $2N$ solutions. Furthermore, let $L$ denote the $2N \times 2N$ Haldane operator matrix. Then we can express (\ref{r_a}) and (\ref{r_b}) as the eigenvalue problem 
%%%%%%%%%%%%%%%%%%%%%%%%%%%%%%%%%%%%%%%%%%%%%%%%%%%%%%%%%%%%%%
   \begin{equation}
       \label{eigProb}
       L (r) {\bf c} = \lambda(r) {\bf c}. 
   \end{equation}
%%%%%%%%%%%%%%%%%%%%%%%%%%%%%%%%%%%%%%%%%%%%%%%%%%%%%%%%%%%%%%
The edge spectral bands are plotted in Fig.~\ref{linearbands}(A) for the parameters corresponding to a topologically nontrivial state. 
Using these parameters over one period in $r$,
we calculate the eigenvalues of  the linear matrix.

These bands %, like in Fig.~\ref{linearbands}(A),  
show that the frequency $\lambda$ depends on the value of $r$ and spectral values in the mid-gap support  localized eigenmodes (edge states). %there is a spectral bands gap where %gapless exist. 
More relevant for below, notice that no bulk modes exists for $-0.5 \le \lambda \le 0.5$, and %but 
the total range of linear spectrum runs from $-3 \le \lambda \le 3$.

  \subsection{Local Chern Number and Topology}
\label{lChern}
  While the Chern number is the classic quantity to probe the topology, it has  limitations. Notably, the Chern number characterizes the topology of bulk (infinite) systems, but can not be directly applied to diagnose genuinely finite systems. Moreover, it is not clear how to apply the Chern number in nonlinear systems which may not posses a spectral band theory. To address this issue we apply the spectral localizer \cite{Cerj2023,Cerjan2024}, a local marker which probes the topology at a user chosen space-time location. 
  A benefit of a local marker approach is that one can probe genuinely finite systems without having to resort to the bulk-edge correspondence.

The spectral localizer is defined by  
  % Let $H(c)=(L+N_m(c))c$ and the define
  \scriptsize
%%%%%%%%%%%%%%%%%%%%%%%%%%%%%%%%%%%%%%%%%%%%%%%%%%%%%%
   \begin{align}
       \label{lLam}
           \mathcal{L}_{(x,y,E)}&(X,Y,H) &\\ \nonumber =
          & \begin{pmatrix}
               (H- E I) && \kappa_x(X-xI)-i\kappa_y(Y-yI)\\\kappa_x(X-xI)+i\kappa_y(Y-yI) && -(H-E I)
           \end{pmatrix}
   \end{align}
%%%%%%%%%%%%%%%%%%%%%%%%%%%%%%%%%%%%%%%%%%%%%%%%%%%%%%
\normalsize
   where $X$ and $Y$ denote diagonal position matrices containing the locations of the lattice sites, and $H$ is the Hamiltonian matrix representing a finite lattice system with open  (zero) boundary conditions. The parameters $\kappa_x,\kappa_y$ are included  to  balance the spatial and energy scales. The probe locations are at $(x,y)$ in space and $E$ in the frequency/energy. The local topology is  classified using the local Chern number index 
%%%%%%%%%%%%%%%%%%%%%%%%%%%%%%%%%%%%%%%%%%%%%%%%%%%%%%
   \begin{equation}
       \label{local_chern}
       C^L=\frac{1}{2}\text{sig}[\mathcal{L}_{(x,y,E)}(X,Y,H)] ,
   \end{equation}
%%%%%%%%%%%%%%%%%%%%%%%%%%%%%%%%%%%%%%%%%%%%%%%%%%%%%%
   where $\text{sig}(M)$ is the signature of the matrix $M$, defined as the number of positive eigenvalues minus the number of negative.

\begin{figure}
\centering
\includegraphics[scale=.225]{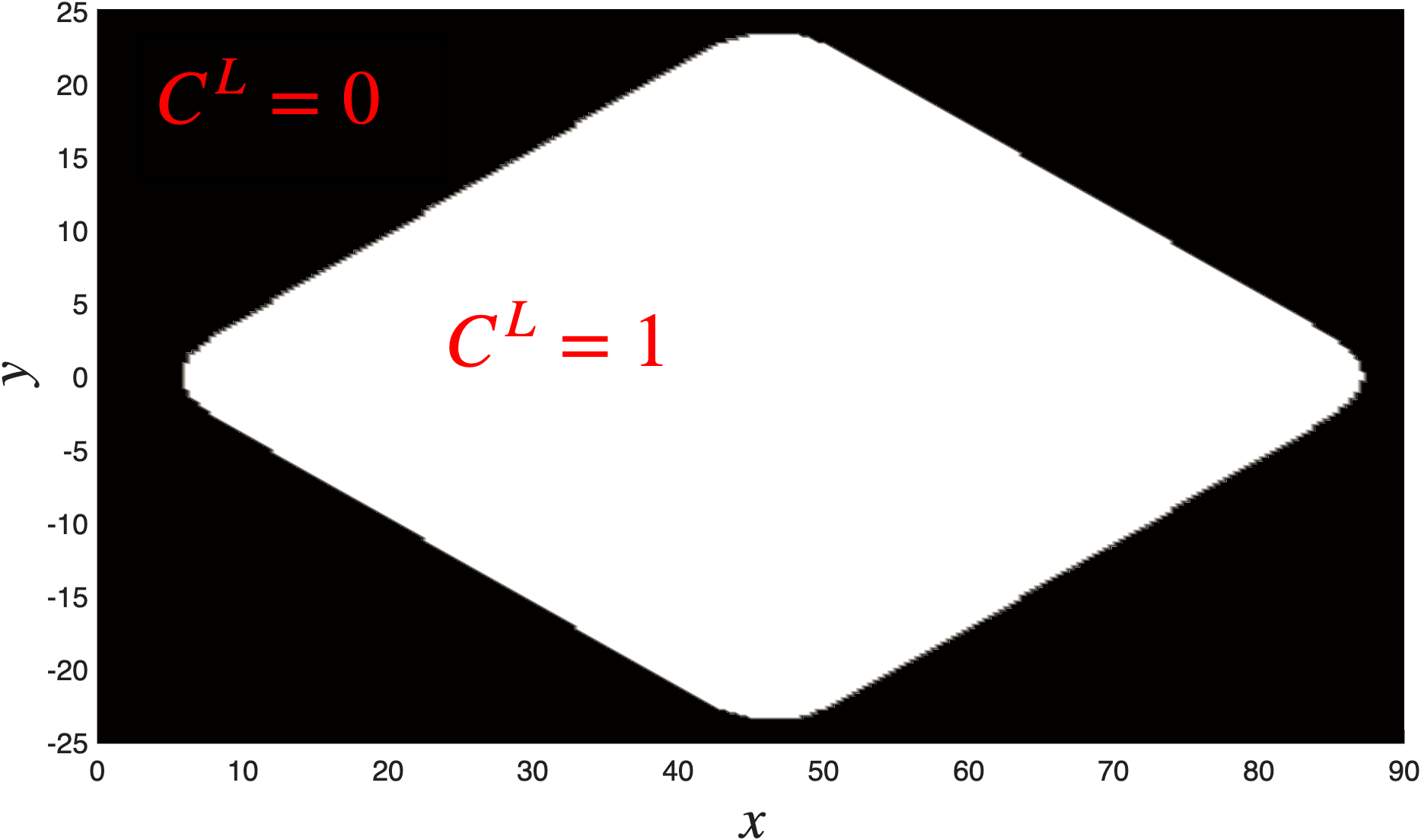}
\caption{Calculation of  the local Chern number in the (linear) Haldane model at $E=0$. The white (black) region corresponds to a local Chern number (\ref{local_chern}) of 1(0). The parameters used are  $N=30$, $\kappa_y=0.1$ and $\kappa_x=\kappa_y/\sqrt{3}$ with
$M=0$,  $t_1=1$, $t_2=0.1$ and $\phi=\frac{\pi}{2}$;  the same as the spectral scan shown in Fig.~\ref{linearbands}. \label{haldaneY}}
\end{figure}

 The local topology of the Haldane model was originally characterized in \cite{Cerjan2022}. However, for convenience we show some localizer outputs for the linear model here. If we fix  frequency at $E = 0$  and scan the spatial points inside and outside of the lattice, the local Chern values are given in Fig.~\ref{haldaneY}. As expected, there is a nonzero topology inside the lattice region and a trivial topology outside. 

 Next, the topology for different values of $E$ is considered. When we fix $y = 0$ and do a scan in $(x,E)$, we obtain the region of topology shown in Fig.~\ref{linearbands}(B). Inside the lattice we observe nontrivial topology for $-1 \le E \le 1$ (in between green lines). These values coincide  with the spectral band gaps shown in Fig.~\ref{linearbands}(A) and (unsurprisingly) predict nontrivial topology in the mid-gap spectral region. This indicates that the local Chern number, unlike the traditional Chern number, is able to {\it directly} identify gaps supporting topologically protected chiral modes. On the other hand, the local Chern number confirms the bulk-edge correspondence. Namely, that the gapless edge modes correspond to a nonzero Chern topology.

While the spectral localizer was originally derived for linear systems, it has recently been extended to nonlinear problems  \cite{Cerj2023}. This is a useful tool as it allows us to probe the local topology for the nonlinear models defined next. %of nonlinear systems as well.

%%%%%%%%%%%%%%
%%%%%%%%%%%%%%%%%%%%%%%%%%%%

%%%%%%%%%%%%%%%%%%%%%%%%%%%%%%%%%%%%%%%%%%%%%%%%

%%%%%%%%%%%%%%%%%%%%%%%%%%%%%%%%%%%%%%%%%%%%%%%%
\section{Some Nonlinear Haldane Systems}
\label{NL_Haldane_Sec}
%%%%%%%%%%%%%%%%%%%%%%%%%%%%%%%%%%%%%%%%%%%%%%%%

In this section we introduce and compare two different nonlinear
 Haldane models. %The comparison highlights the stark contrast in their dynamics. 
 Both are modifications of the linear system given in  Eq.~(\ref{ab_sites}).
 
 The first model we consider is an on-site (local) Kerr-type given by
 %%%%%%%%%%%%%%%%%%%%%%%%%%%%%%%%%%%%%%%%%%%%%%%%%%%%%%%%%%%%%%%%%%%%%%
   \begin{equation}
        \label{nonlin_ab}
        \begin{split}
       i\frac{da_{mn}}{dt}=H_a[a_{mn},b_{mn}]+ \sigma|a_{mn}|^2 a_{mn} ,\\
        i\frac{db_{mn}}{dt}=H_b[a_{mn},b_{mn}]+\sigma|b_{mn}|^2b_{mn} .
        \end{split}
   \end{equation}
%%%%%%%%%%%%%%%%%%%%%%%%%%%%%%%%%%%%%%%%%%%%%%%%%%%%%%%%%%%%%%%%%%%%%%
which we refer to as the nonlinear Haldane (NH) model. %Here $\sigma \in \RR$ denotes the strength and type (focusing or defocusing) of the nonlinearity. 
Here $\sigma \geq0$ denotes the strength of the focusing  nonlinearity. This system reflects a natural tight-binding approximation of a weakly nonlinear system on a honeycomb lattice \cite{AblowitzReview}. It can be viewed as a natural 2D extension of the discrete NLS equation \cite{Kevrekidis2009,Christodoulides1988}.
The second model we examine is the (nonlocal) Ablowitz-Ladik-type system
%%%%%%%%%%%%%%%%%%%%%%%%%%%%%%%%%%%%%%%%%%%%%%%%%%%%%%%%%%%%%%%%%%%%%
  \begin{equation}
       \label{alhnonlin_ab}
       \begin{split}
       i\frac{da_{mn}}{dt}=(1+\sigma|a_{mn}|^2)H_a[a_{mn},b_{mn}] ,\\  i\frac{db_{mn}}{dt}=(1+\sigma|b_{mn}|^2) H_b[a_{mn},b_{mn}] .
       \end{split}
   \end{equation}  
%%%%%%%%%%%%%%%%%%%%%%%%%%%%%%%%%%%%%%%%%%%%%%%%%%%%%%%%%%%%%%%%%%%%%
which we refer to %it
as the Ablowitz-Ladik-Haldane (ALH) model. 
Once again, $\sigma \geq 0$ denotes the strength of the focusing nonlinearity. This system is a natural extension of the classic scalar AL model, which is an integrable discrete nonlinear Schr\"odinger-type equation \cite{Ablowitz1975,Ablowitz1976}. A 1D version of Eq.~(\ref{alhnonlin_ab}) was examined in \cite{Yang2002}. Unlike the nonlocal nonlinear model considered in \cite{Cerj2023}, the intensity contribution in Eq.~(\ref{alhnonlin_ab}) is local.

The main difference between these two models is which terms experience a nonlinear intensity effect. In the NH model there is only an on-site self-interaction contribution, while in the %NLH 
ALH model, all nearest  and next-nearest neighbors are affected by the intensity. For example, the nonlinear nearest neighbor terms in first ALH equation read $t_1 |a_{mn}|^2 (b_{mn} + b_{m-1,n} + b_{m,n-1})$, indicating the intensity of $a_{mn}$ modulates the three nearest neighbors. 

Physically, the NH model is easier to justify in a tight-binding approximation since most nonlinear effects are rather weak \cite{AblowitzReview}. However, even in 1D, the  discrete NLS equation, the natural analog of the NH model,  struggles to  support simple traveling soliton solutions due to the  Peierls-Nabarro effect \cite{Jenkinson2015,Flach1999}. On the other hand, the 1D AL model is integrable and possess a systematic way of obtaining traveling soliton solutions and an infinite number of conservation laws \cite{Ablowitz2004}. Several recent works have explored AL models in the context of 1D topological insulators \cite{Munoz2017,Castro2025}.
%This  observation is the motivation to seek an AL-type equation in 2D honeycomb  systems. 

An important feature of most physical models in closed systems is the presence of conserved quantities.
The NH model  preserves the dynamical invariant power, given by %mass, 
%%%%%%%%%%%%%%%%%%%%%%%%%%%%%%%%%%%%%%%%%%%%%%%%%%%%%%%%%%%%%%%%
   \begin{equation}
       \label{nhmass}
       \mathbb{P}_{\text{NH}}=\sum_{m,n} |a_{mn}|^2+|b_{mn}|^2 ,
   \end{equation}
%%%%%%%%%%%%%%%%%%%%%%%%%%%%%%%%%%%%%%%%%%%%%%%%%%%%%%%%%%%%%%%%
for zero boundary conditions. On the other hand, the dynamical invariant for total power of the ALH model is given by 
%%%%%%%%%%%%%%%%%%%%%%%%%%%%%%%%%%%%%%%%%%%%%%%%%%%%%%%%%%%%%%%%
   \begin{equation}
       \label{alh_mass}
       \mathbb{P}_{\text{ALH}}=\sum_{m,n} \log(1+\sigma|a_{mn}|^2)+\log(1+\sigma|b_{mn}|^2) ,
   \end{equation}
%%%%%%%%%%%%%%%%%%%%%%%%%%%%%%%%%%%%%%%%%%%%%%%%%%%%%%%%%%%%%%%%
for zero boundary conditions. Clearly, this is not the standard power defined in terms of the $\ell_2$ norm. However, AL models typically involve local  densities  with logarithms \cite{Ablowitz2004}. Notice in the weak nonlinear limit that $\log (1 + \sigma |a_{mn}|^2) \approx \sigma |a_{mn}|^2$ as $|a_{mn}|^2 \rightarrow 0$ (and similarly for $b_{mn}$).

Both the NH and ALH model also possess a conservation of energy 
%%%%%%%%%%%%%%%%%%%%%%%%%%%%%%%%
 \begin{equation}
       \label{nh_ham}
       \begin{split}
       \mathbb{H}_{\text{NH}}=\sum_{m,n} t_2\Big(a_{mn}^*[e^{i\phi}(a_{m,n+1}+a_{m-1,n}+a_{m+1,n-1})\\+e^{-i\phi}(a_{m+1,n}+a_{m,n-1}+a_{m-1,n+1})]\\+ b_{mn}^*[e^{-i\phi}(b_{m,n+1}+b_{m-1,n}+b_{m+1,n-1})\\+e^{i\phi}(b_{m+1,n}+b_{m,n-1}+b_{m-1,n+1})]\Big)\\
       +t_1\Big( a_{mn}^*[ b_{mn}+b_{m-1,n}+b_{m,n-1}] \\+b_{mn}^*(a_{mn}+a_{m+1,n}+a_{m,n+1})\Big) \\
      + \frac{\sigma}{2} |a_{mn}|^4 +  \frac{\sigma}{2} |b_{mn}|^4 \\+M\Big(|a_{mn}|^2-|b_{mn}|^2\Big)
       \end{split}    
   \end{equation}
%%%%%%%%%%%%%%%%%%%%%%%%%%%%%%%%
 and
%%%%%%%%%%%%%%%%%%%%%%%%%%%%%%%%   
 \begin{equation}
       \label{alh_ham}
       \begin{split}
       \mathbb{H}_{\text{ALH}}= \sum_{m,n} t_2\Big(a_{mn}^*[e^{i\phi}(a_{m,n+1}+a_{m-1,n}+a_{m+1,n-1})]\\+e^{-i\phi}(a_{m+1,n}+a_{m,n-1}+a_{m-1,n+1})]\\+ b_{mn}^*[e^{-i\phi}(b_{m,n+1}+b_{m-1,n}+b_{m+1,n-1})\\+e^{i\phi}(b_{m+1,n}+b_{m,n-1}+b_{m-1,n+1})]\Big)\\
       +t_1\Big( a_{mn}^*[ b_{mn}+b_{m-1,n}+b_{m,n-1}] \\+b_{mn}^*(a_{mn}+a_{m+1,n}+a_{m,n+1})\Big)\\ +M\Big(|a_{mn}|^2-|b_{mn}|^2\Big) 
       \end{split}    
   \end{equation}
%%%%%%%%%%%%%%%%%%%%%%%%%%%%%%%%%%%%
for zero boundary conditions, respectively.
The NH governing equations (\ref{nonlin_ab}) form a Hamiltonian system and are recovered from the energy in Eq.~(\ref{nh_ham}) by 
%%%%%%%%%%%%%%%%%%%%%%%%%%%%%%%%%%%%%%%%%%%%%%%%%%%%%%%%%%%%%%%%%%%%%
  \begin{equation}
       \label{nhnonlin_ab_hamil}
       i\frac{da_{mn}}{dt}= \frac{\partial \mathbb{H}_\text{NH}}{ \partial a_{mn}^*} , ~~~~~      i\frac{db_{mn}}{dt}= \frac{\partial \mathbb{H}_\text{NH}}{ \partial b_{mn}^*} ,
   \end{equation}  
%%%%%%%%%%%%%%%%%%%%%%%%%%%%%%%%%%%%%%%%%%%%%%%%%%%%%%%%%%%%%%%%%%%%%
%where $\delta/\delta c$ denotes the variational derivative with respect to the function $c$. On the other hand, the ALH model requires a modified definition since it does not depend explicitly on the nonlinearity. The ALH governing equations (\ref{alhnonlin_ab}) form a Hamiltonian system
On the other hand, the ALH model requires a modified definition since it does not depend explicitly on the nonlinearity. The ALH governing equations (\ref{alhnonlin_ab}) form a Hamiltonian system
%%%%%%%%%%%%%%%%%%%%%%%%%%%%%%%%%%%%%%%%%%%%%%%%%%%%%%%%%%%%%%%%%%%%%
  \begin{equation}
       \label{alhnonlin_ab_hamil}
       \begin{split}
       i\frac{da_{mn}}{dt}= (1+ \sigma |a_{mn}|^2)\frac{\partial \mathbb{H}_\text{ALH}}{ \partial a_{mn}^*} , \\
       i\frac{db_{mn}}{dt}= (1 + \sigma |b_{mn}|^2)  \frac{\partial \mathbb{H}_\text{ALH}}{\partial b_{mn}^*} ,
       \end{split}
   \end{equation}  
%%%%%%%%%%%%%%%%%%%%%%%%%%%%%%%%%%%%%%%%%%%%%%%%%%%%%%%%%%%%%%%%%%%%%
where a nonlinear correction factor is included, mirroring the representation in Eq.~(\ref{alhnonlin_ab}).

\section{Nonlinear edge current}
\label{Travel_Sec}
%%%%%%%%%%%%%%%%%%%%%%%%%%%%%%%%%%%%%%%%%%%%%%%%%%%%%%%   
A quintessential feature of linear Chern insulator systems is the existence of traveling edge currents on the interface of two topologically dissimilar media. These states are chiral and propagate robustly in spite of defects and perturbations. The ability of edge currents  to propagate in  nonlinear models introduced above will be examined in this section. The contrast is stark. 

For all cases considered below, we take a finite-sized lattice with sites at $a_{mn},b_{mn}$ for $m ,n = 1,\dots,N$ for $N = 30$. 
To avoid hanging corners, the sites at the corners, $a_{1,1}$ and $b_{30,30}$, are removed. This slight modification permits a more direct conduit with little effect on the overall form of the edge state.  For all cases considered below, we fix the standard parameters as $\sigma=1$, $M=0$,  $t_1=1$, $t_2=0.1$ and $\phi=\frac{\pi}{2}$. In the linear Haldane model (\ref{ab_sites}), these values correspond to a nontrivial Chern insulator. We explore the $M \not= 0 $ case later. 

To probe the dynamics we take  a simple initial condition and evolve it using a standard 4th-order Runge-Kutta integrator with timestep $\Delta t=10^{-3}$.   The initial condition used is the 1D linear edge solution of Eq.~(\ref{r_a})-(\ref{r_b}) at $\lambda = 0$. This is chosen because: (a) it is simple to compute and (b) it should correspond to a %n eigen
mode in the %topological nontrivial
mid-gap region. Since Eq.~(\ref{r_a})-(\ref{r_b}) are not solved with open boundary conditions in the direction parallel to the edge, embedding it in a 2D problem with zero boundary conditions will result in some radiation at the corners due to the boundary condition mismatch. In our experience, these results appear to be fairly typical and representative of the system dynamics.

%%%%%%%%%%%%%%%%%%%%%%%%%%%%%%%%%%%%%%%%%%%%%%%%%%%%%%%%%%%%%%%%% 

%%%%%%%%%%%%%%%%%%%%%%%%%%%%%%%%%%%%%%%%%%%%%%%%%%%%%%%%%%%%%%%%% 
   \subsection{Traveling wave dynamics in the NH model}
   \label{travNH}
%%%%%%%%%%%%%%%%%%%%%%%%%%%%%%%%%%%%%%%%%%%%%%%%%%%%%%%%%%%%%%%%%

\begin{figure*}
\centering
\includegraphics[scale=.40]{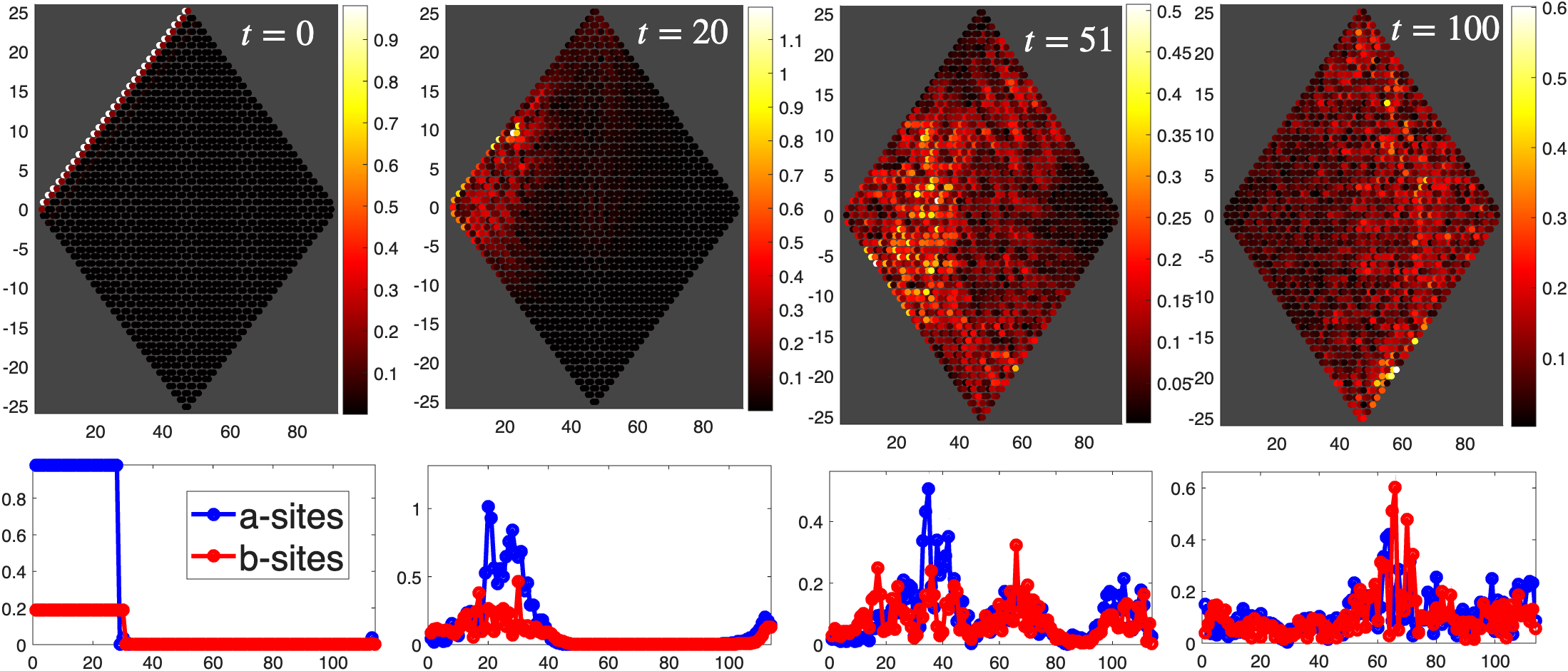}
\caption{(Top row) Evolution of $|a_{mn}|,|b_{mn}|$ in the strongly nonlinear ($\sigma = 1$) NH model (\ref{nonlin_ab}) with %\jtc{the} 
a linear ($\lambda = 0$) edge mode initially embedded at $n =0$. The edge mode  radiates significant amounts of energy. 
 (Bottom row) Edge %Side 
 profile snapshots of the evolution shown in the top row. The $x$-axis is simply indexing each of the 116 edge sites. These magnitudes correspond to the outermost lattice sites, along the perimeter of the domain.  \label{nhNoMovie}}
\end{figure*}

\begin{figure}
   \centering
\includegraphics[scale=.25]{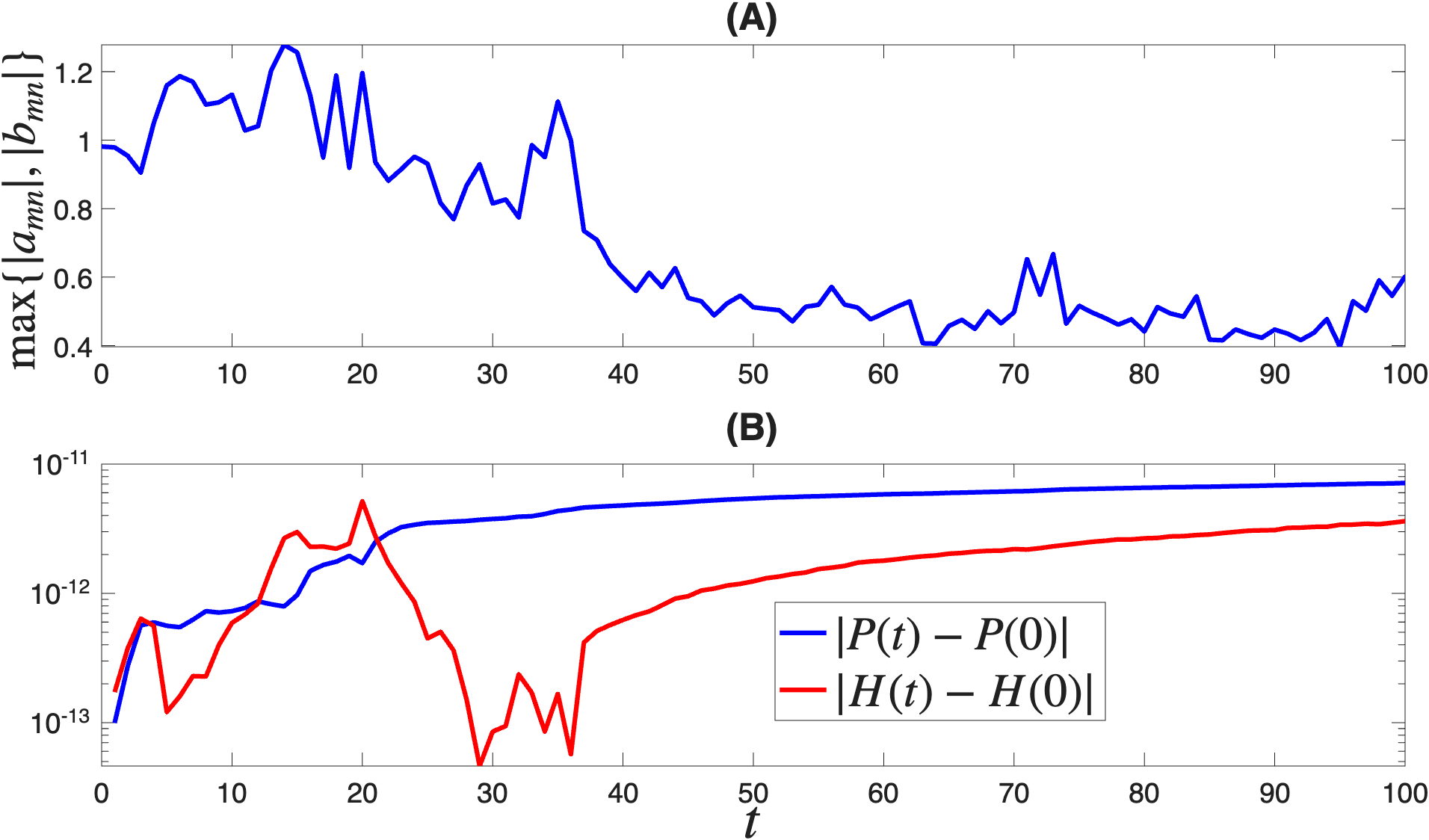}
\caption{(A) Maximum magnitude of the NH solution in Fig.~\ref{nhNoMovie}. (B)  Absolute difference of numerically computed power (energy) in blue (red)  (see Eqs.~(\ref{nhmass}) and (\ref{nh_ham})) at time $t$  and the initial power (energy). \label{nhNoMass}}
\end{figure}

The evolution of a highly nonlinear mode in the NH model is shown in Fig.~\ref{nhNoMovie}. From the initial time, significant amounts of radiation are observed until a coherent structure is no longer discernible. Tracking the peak amplitude in Fig.~\ref{nhNoMass}(A), we observe the magnitude decreases until all but a sea of radiation appears to exist. 
While the localized state disintegrates,
there is no loss of power or energy (see Fig.~\ref{nhNoMass}(B)); this is a conserved quantity up to numerical integration %round-off 
error.

This lack  of coherent evolution can be attributed to the effect of strong on-site nonlinearity. When a strong difference between $|a_{mn}|^2$ and $|b_{mn}|^2$ exists, this acts like a mass term ($M \not= 0$) in the linear Haldane model. When $|M| > 3 \sqrt{3} t_2 |\sin \phi|$ the linear system is a  trivial insulator and  lacks stable chiral states. Another point-of-view is that these highly nonlinear states suffer from  Peierls-Nabarro energy barrier effects. A lack of translational invariance results in a position-dependence of the local energy, typically  %which is 
accommodated by a change such as loss of velocity \cite{Jenkinson2015} or amplitude \cite{Ablowitz2021}.

On the other hand, in \cite{Ablowitz2014,Ablowitz2021} an asymptotic theory was derived for (stable) soliton solutions. In brief, those results %that result 
said that for profiles where the spatial dispersion balances weak self-focusing nonlinearity, to leading-order, edge solitons exist. In \cite{Ablowitz24} a similar nonlinear Haldane model found that for localized profiles in this regime,  robust solitonic behavior can be observed. 

To reconcile these two results, we note that one (the former) concerns strong nonlinearity and the other (the latter) concerns weak nonlinearity. To confirm this we have included  Appendix \ref{weak_sect} which shows  improved edge soliton conduction for the NH model with weak nonlinearity. The results in \cite{Ablowitz2021} indicate that balanced edge solitons can be observed in these types of models after a redistribution of energy, i.e. radiation.

%%%%%%%%%%%%%%%%%%%%%%%%%%%%%%%%%%%%%%%%%%%%%%%%%%%%%%%%%%%%%%%%%%%%%%
%%%%%%%%%%%%%%%%%%%%%%%%%%%%%%%%%%%%%%%%%%%%%%%%%%%%%%%%%%%%%%%%%%%%%%
%%%%%%%%%%%%%%%%%%%%%%%%%%%%%%%%%%%%%%%%%%%%%%%%%%%%%%%%%%%%%%%%%%%%%%

\begin{figure*}
\centering
\includegraphics[scale=.40]{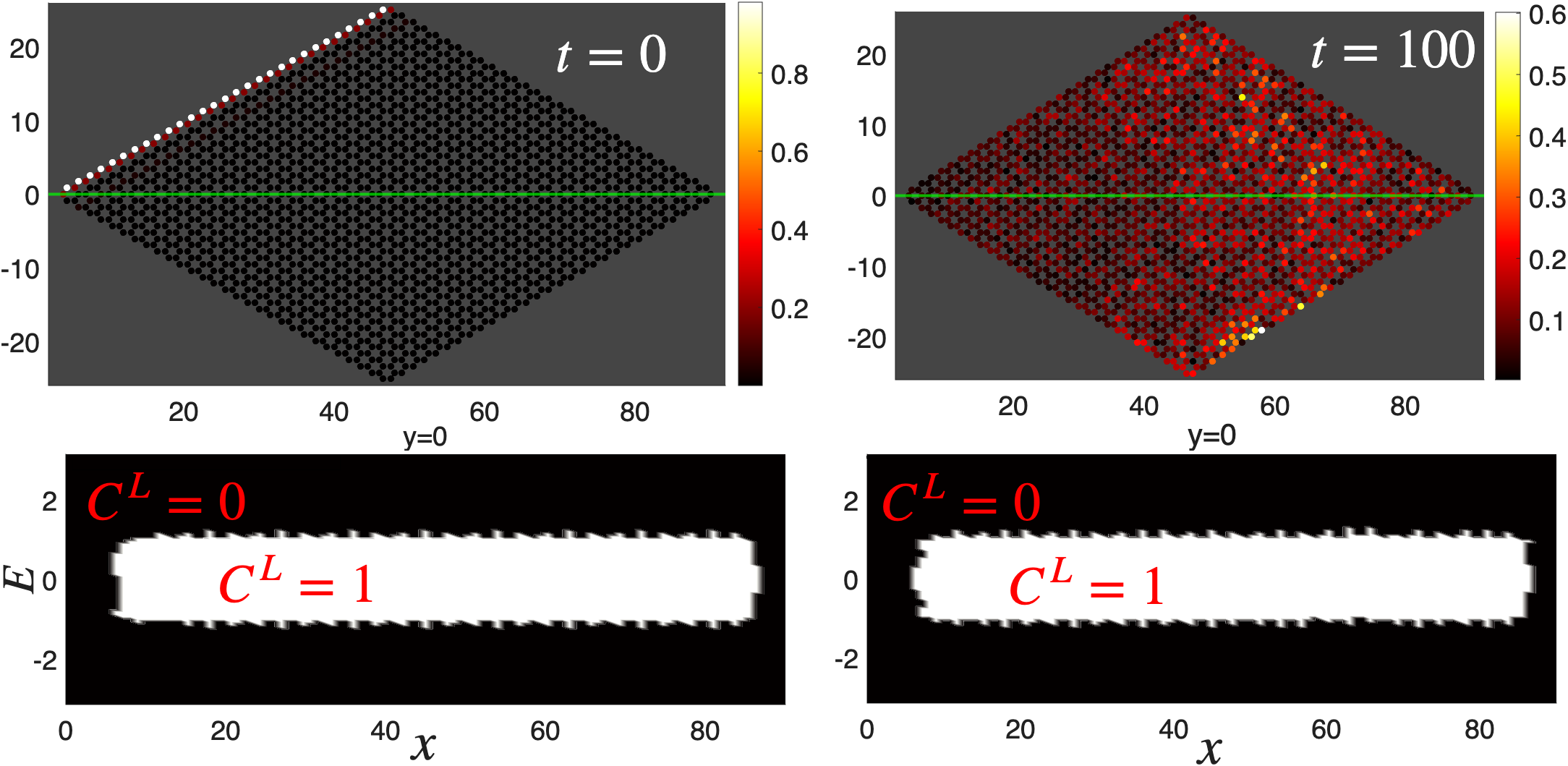}
\caption{Calculation of the local Chern number (\ref{local_chern}) at $y=0$ (green line in top row) and various $E$ values of the NH model (\ref{nonlin_ab}) for the initial condition used in Fig.~\ref{nhNoMovie} with standard parameters. }  \label{NHtop}
\end{figure*}

A natural question to ask is if these modes are topological. To address this question, we apply the spectral localizer in Eq.~(\ref{lLam}) using  a nonlinear Hamiltonian. The effective Hamiltonian used is $H_\text{lin} + N[a_{mn},b_{mn}]$, where $H_\text{lin}$ is the linear  Haldane model and 
$N[a_{mn},b_{mn}] = \text{diag}(|a_{mn}|^2, |b_{mn}|^2 )$ captures the onsite nonlinearity. 
Since this is not a stationary solution, we probe the solution 
at different times and compute its local topology. The  local Chern number at times $t=0$ and 
$t=100$ for points in the middle of the lattice at $y=0$ is shown in Fig.~\ref{NHtop} for  various $x$ and $E$; these results are typical. We note  that this is almost identical to the topology results found in the linear case. 

\begin{figure}
   \centering
\includegraphics[scale=.225]{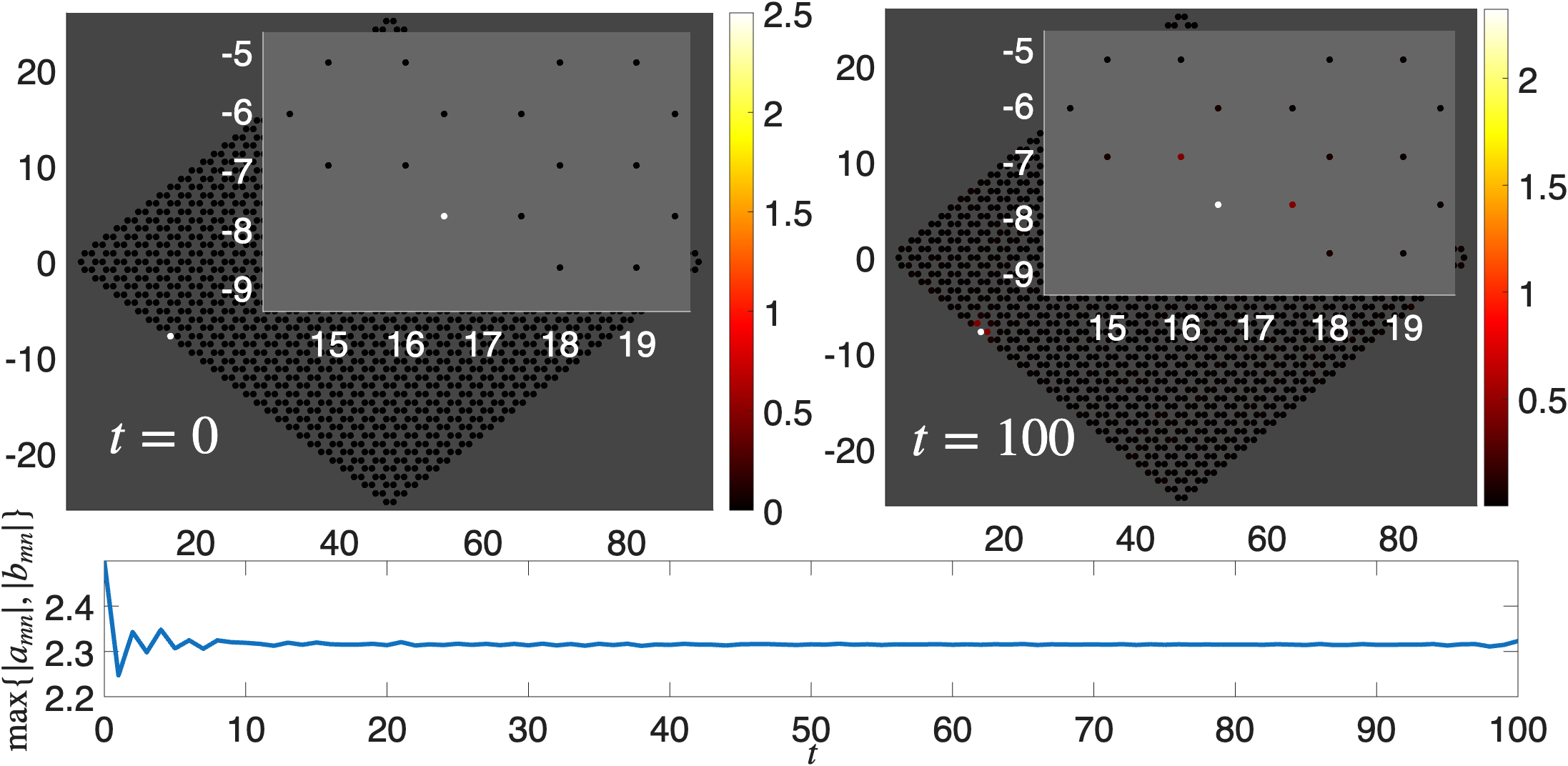}
\caption{ (Top row) Evolution of $|a_{mn}|,|b_{mn}|$ in the strongly nonlinear ($\sigma = 1$) NH model (\ref{nonlin_ab}) with 
an initial condition of %$a_{1,10}=5$ and $b_{1,10}=1$. 
$a_{1,10}=2.5$. The mode remains stationary with a slight variation of amplitude. (Bottom row) Tracking the max amplitude of the mode as a function of time } \label{nhGen}
\end{figure}

On the other hand, consider an initial condition which is very localized %(excite only two 
(excite only one site) and %large 
moderate amplitude ($a_{1,10}=2.5$) %and $b_{1,10}=1$). 
In this case, we excite a  solitary mode which maintains a stable but mostly fixed profile; see Fig.~\ref{nhGen}. This is quite different from the previous case. The reason this mode is not traveling is because it corresponds to a stationary gap soliton mode, of the sort discussed in \cite{Chong2016} and Appendix~\ref{stationary_ALH_sec}. 

To confirm this, we calculate the Rayleigh quotient
%%%%%%%%%%%%%%%%%%%%%%%%%%%
\begin{equation}
\label{RQ}
R({\bf x}) = \frac{ \langle {\bf x} | H_{\rm lin} + N |{\bf x} \rangle}{ || {\bf x} ||^2 } ,
\end{equation}
%%%%%%%%%%%%%%%%%%%%%%%%%%%
which when ${\bf x}$ is an eigenmode of the operator $H_{\rm lin} + N$, returns the corresponding eigenvalue. For the example shown in Fig.~\ref{nhGen}, we find $R({\bf x}) \approx 5.43$ which, for a single eigenmode, clearly lies {\it above} the linear band region, $- 1\le \lambda \le 1 $ and thus corresponds to a null local Chern number.

%To further investigate this phenomenon, we repeat our numerics with an even simpler initial condition; that is taking an $a$-site and corresponding $b$-site on the edge of the lattice and setting each of them to a positive real number. All other sites are $0$. Fig.~\ref{nhGen} shows typical results of this method. In this case we set $a_{1,10}=5$ and $b_{1,10}=1$. The mode remains almost entirely stationary with some slight variation in the amplitude.  

How does one reconcile these findings? A {\it nontrivial} local Chern number indicates a necessary but not sufficient condition for stable traveling edge solitons. Other considerations, such as the Peierls-Nabarro energy barrier, also play a role when evolving.This appears to be an open area of research: the role stability (or instability) plays in the evolution of %in 
nonlinear topological insulators. A thorough stability analysis of these nonlinear modes is outside the scope of this work.  However, for all cases we examined with a {\it trivial} local Chern number, we do not observe robust chiral solitons, for both the NH and subsequent ALH models.

%How does one reconcile this finding? In the case of nonlinearity, there are two notions: topology and stability. While a nontrivial local Chern number indicates a chiral  flow of energy, %nature to energy flow, 
%highly nonlinear solitary structures can be % are 
%unstable and rapidly disintegrate upon impact with corners. This is quite different from the linear system. 

%%%%%%%%%%%%%%
%%%%%%%%%%%%%%%%%%%%%%%%%%%%

%%%%%%%%%%%%%%%%%%%%%%%%%%%%%%%%%%%%%%%%%%%%%%%%%%%%%%%%%%%%%%%%%%%%%%
%%%%%%%%%%%%%%%%%%%%%%%%%%%%%%%%%%%%%%%%%%%%%%%%%%%%%%%%%%%%%%%%%%%%%%

%%%%%%%%%%%%%%%%%%%%%%%%%%%%%%%%%%%%%%%%%%%%%%%%%%%%%%%%%%%%%%%%%%%
   \subsection{Traveling solitary waves in the ALH model}
   \label{alh_sect}
%%%%%%%%%%%%%%%%%%%%%%%%%%%%%%%%%%%%%%%%%%%%%%%%%%%%%%%%%%%%%%%%%%%

\begin{figure*}
   \centering
\includegraphics[scale=.40]{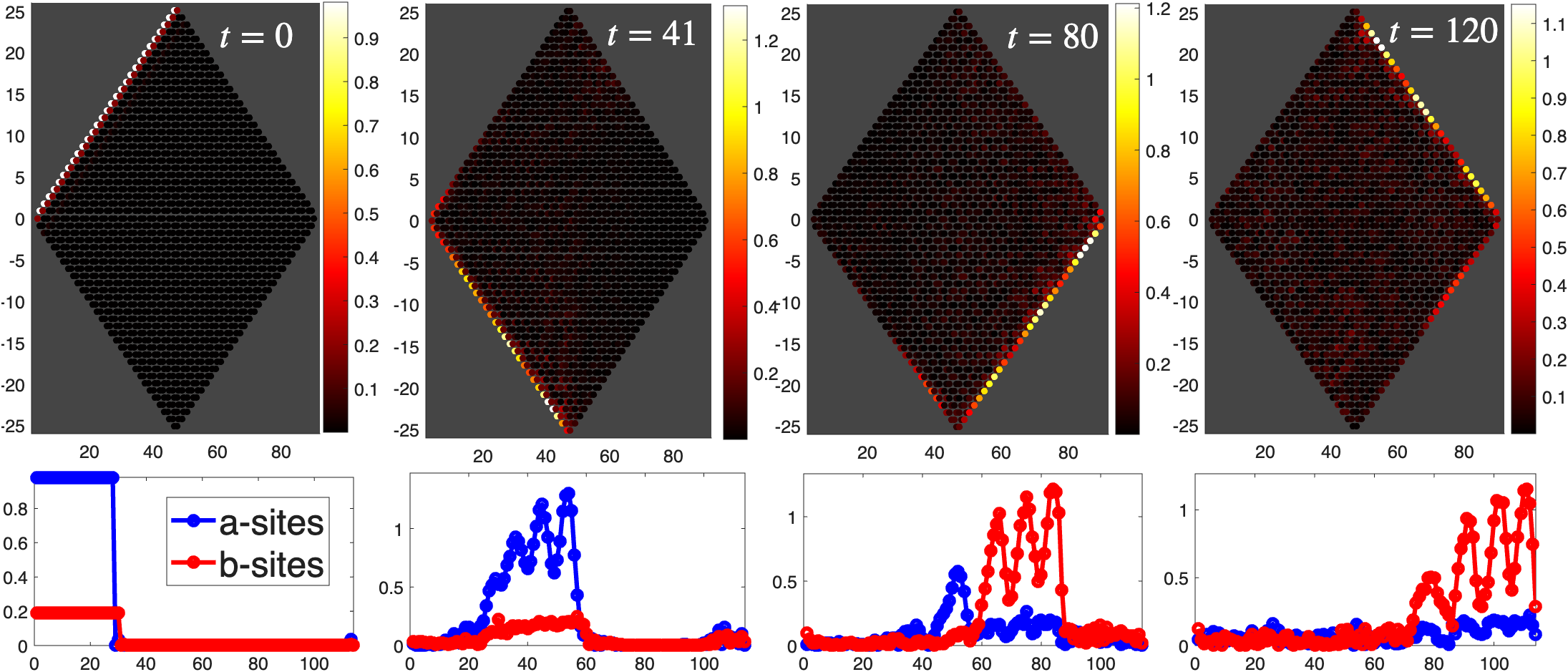}
\caption{(Top row) Evolution of $|a_{mn}|,|b_{mn}|$ in the strongly nonlinear ($\sigma = 1$) ALH model (\ref{alhnonlin_ab}) with a 
linear ($\lambda = 0$) edge mode initially embedded at $n =0$.  Relative to Fig.~\ref{nhNoMovie}, the edge mode does not emit much radiation. 
 (Bottom row) Edge 
 profile snapshots of the evolution shown in the top row. These magnitudes correspond to the outermost lattice sites, along the perimeter of the domain. One can observe the formation of a solitonic wave train.  \label{alhNoMovie}}
   \end{figure*}

\begin{figure}
   \centering
\includegraphics[scale=.23]{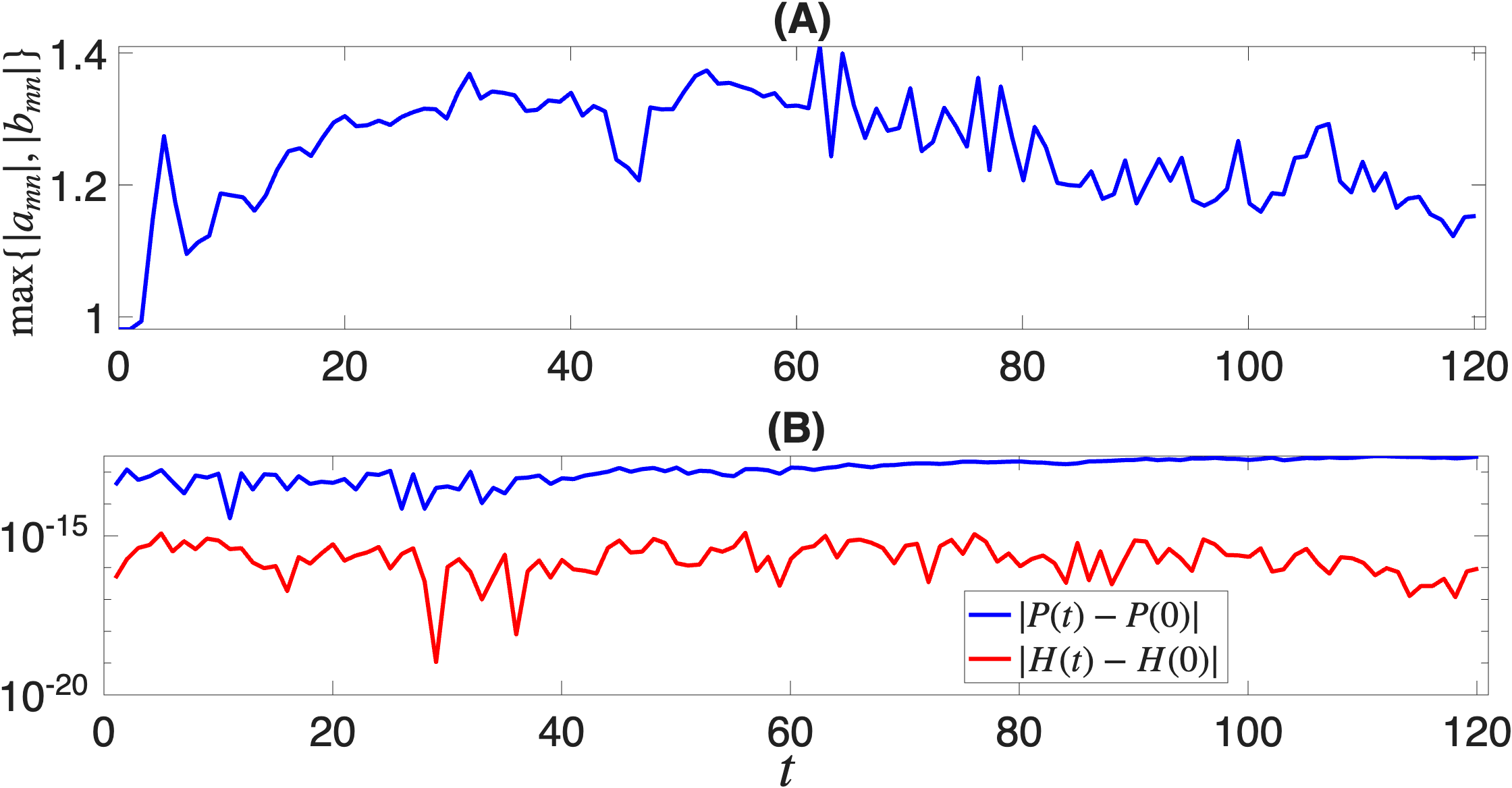}
\caption{(A) Maximum magnitude of the ALH solution in Fig.~\ref{alhNoMovie}. (B)  Absolute difference of numerically computed power (energy) in blue (red) (see Eqs.~(\ref{alh_mass}) and (\ref{alh_ham})) at time $t$  and the initial power (energy).} \label{alhNoMass}
\end{figure}

We now turn our attention to the ALH model in Eq.~(\ref{alhnonlin_ab}). To begin, we repeat the experiment conducted for the NH equation in Fig.~\ref{nhNoMovie}. That is, we take the same initial condition and standard parameters for the linear 
part of the equation; the evolution of the ALH model for relatively strong nonlinearity is summarized in Fig.~\ref{alhNoMovie}. There is a striking difference between these two models. The edge current in the ALH case remains coherent, even at these rather large powers, and propagates counterclockwise around the lattice. Similar to the NH model, the ALH is observed to conserve power and energy, up to numerical integration %round-off 
error (see Fig.~\ref{alhNoMass}(B))

Unlike the NH case, the peak magnitude shown in Fig.~\ref{alhNoMass}(A) does not decrease over the time scales shown. Rather, after sufficiently long time, we observe the formation of a wave train of solitary waves. This is highlighted in the bottom row of Fig.~\ref{alhNoMovie} which shows the solution magnitude along the boundary; the edge mode profile. For example, by time $t = 120$ we can identify four distinct soliton-like peaks that propagate at nearly the same velocity. These solitary waves were observed to remain distinct as they propagate along the edge for long times. 
%We refer to this as {\it soliton fission} (dissolution into smaller solitary waves).
A wave train was also formed in the context of (time-reversal variant) atomic vapors \cite{Zhang2020}, however those edge solitons propagate bi-directionally. That we observe this from these initial conditions speaks to their robustness and persistence.

%%%%%%%%%%%%%%%%%%%%%%%%%%%%%%%%%%%%%%%%%%%%%%%%%%%%%%%%%%%%%%%%%%%
%%%%%%%%%%%%%%%%%%%%%%%%%%%%%%%%%%%%%%%%%%%%%%%%%%%%%%%%%%%%%%%%%%%
%%%%%%%%%%%%%%%%%%%%%%%%%%%%%%%%%%%%%%%%%%%%%%%%%%%%%%%%%%%%%%%%%%%

%%%%%%%%%%%%%%%
   \begin{figure*}
   \centering
\includegraphics[scale=.4]{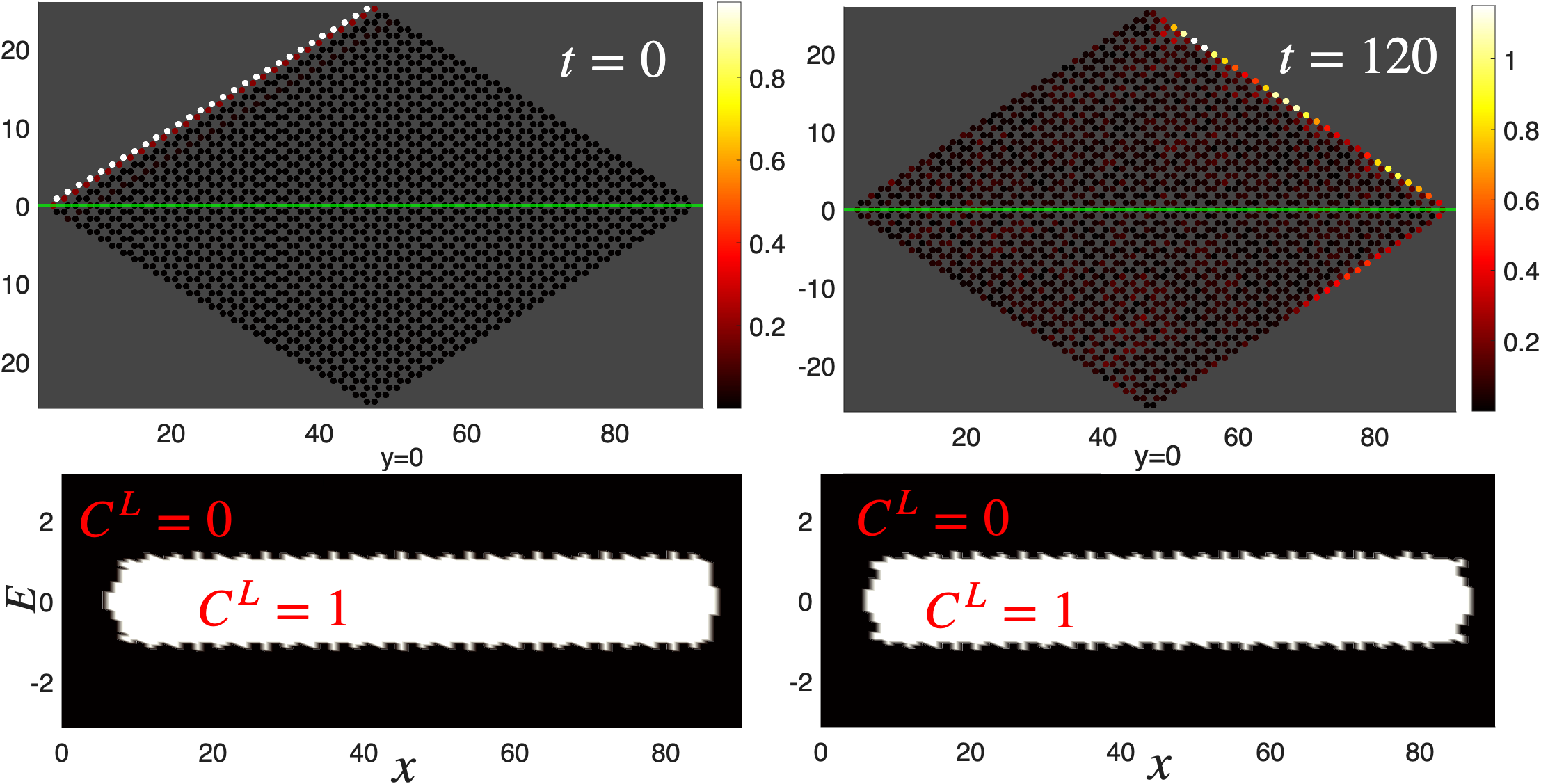}
\caption{Calculation of the local Chern number (\ref{local_chern}) at $y=0$ (green line in top row) and various $E$ values of the ALH model (\ref{alhnonlin_ab}) for the initial condition used in Fig.~\ref{alhNoMovie} with standard parameters.} \label{alhNoTop}
   \end{figure*}
%%%%%%%%%%%%%%%%%%%%%%%%%%%%%%%%%%%

The (local) topology of this system was %is 
computed and is shown in Fig.~\ref{alhNoTop}. Similar to the linear and NH cases for these parameters, there is nonzero local Chern number  in the interior of the lattice and in the frequencies corresponding to the middle linear band gap. We have consistently observed robust traveling edge solitons when the local Chern number is nonzero. In our experience, it is a good indicator of chiral solitons in the ALH model.
%As a result, topology here behaves like linear topology for the parameters we consider. On the other hand, clearly this is not enough to guarantee a coherent solitonic structure (recall Fig.~\ref{nhNoMovie}). \tij{TIJ: Do we want to revist this?} More insight is clearly needed, but is outside the scope of this work.

On the other hand, we also explore
%Here we briefly detail 
the effects of setting the mass term, $M$, to be nonzero. We repeat the experiment conducted for the ALH equation in Fig.~\ref{alhNoMovie}, the only differnce being that we set $M=1$ instead of $0$. We note that this $M$ value is larger than $3 \sqrt{3} t_2 |\sin \phi|$, which leads to a trivial insulator in the linear Haldane model.

\begin{figure*}
   \centering
\includegraphics[scale=.40]{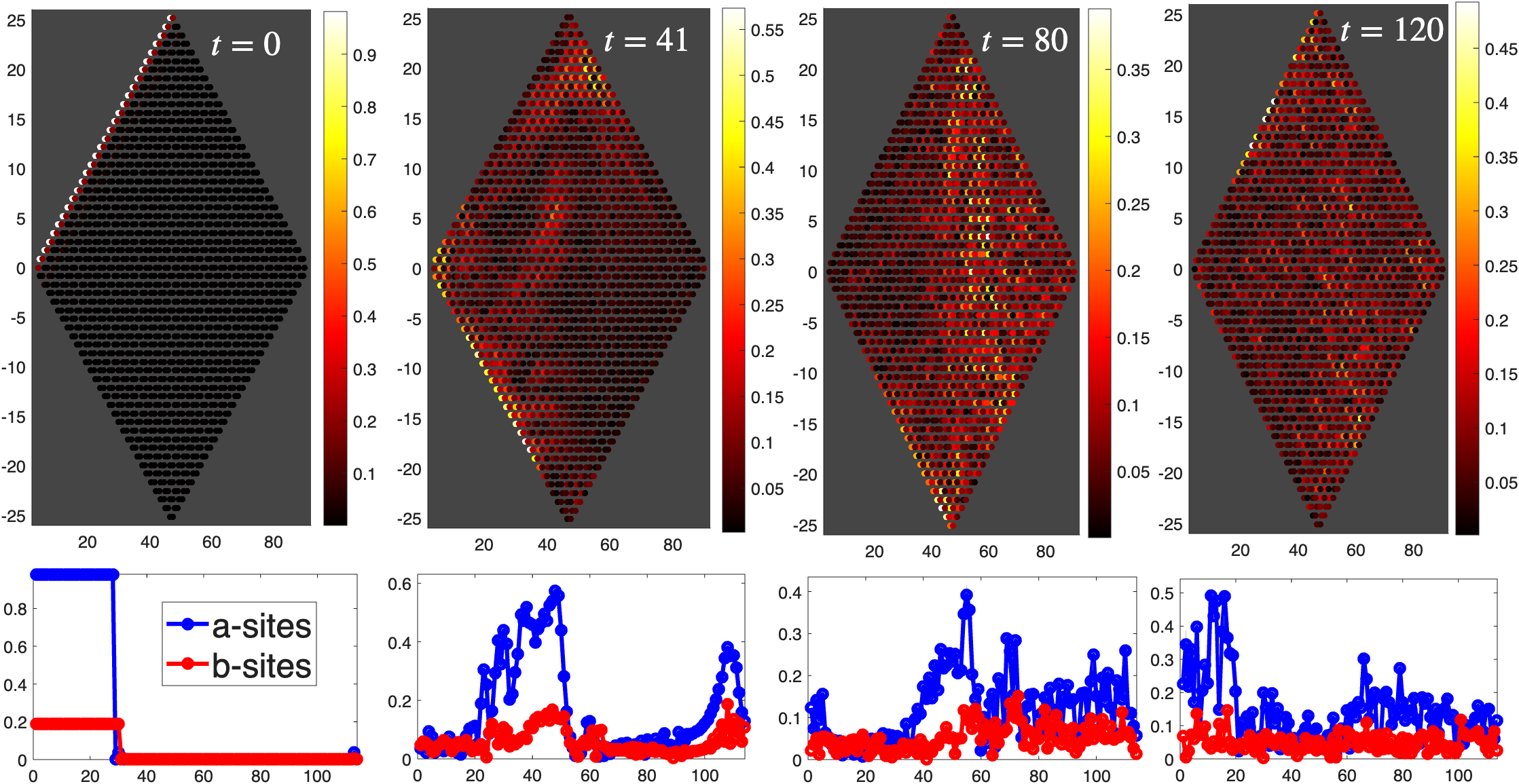}
\caption{(Top row) Evolution of $|a_{mn}|,|b_{mn}|$ in the ALH model (\ref{alhnonlin_ab}) with a 
linear ($\lambda = 0$) edge mode initially embedded at $n =0$. The parameters are the same as used in Fig.~\ref{alhNoMovie}, except here $M=1$ instead of $0$. Relative to Fig.~\ref{alhNoMovie}, the edge mode emits significantly more radiation. (Bottom row) Edge profile snapshots of the evolution shown in the top row. %These magnitudes correspond to the outermost lattice sites, along the perimeter of the domain. 
\label{alh_M1_Movie}}
   \end{figure*}

The evolution is shown in Fig.~\ref{alh_M1_Movie}. There is a noticeable difference between these two cases, with the $M=1$ case radiating significantly more. This is the key reason for us focusing on the case of $M=0$. The local Chern number for this case is zero near the $E = 0$ energy values; hence these results are not unexpected.

%%%%%%%%%%%%%%%%%%%%%%%%%%%%%%%%%%%%%%%%%%%%%%%%%%%%%%%%%%%%%%%%%%%
%%%%%%%%%%%%%%%%%%%%%%%%%%%%%%%%%%%%%%%%%%%%%%%%%%%%%%%%%%%%%%%%%%%

\begin{figure*}
   \centering
\includegraphics[scale=.4]{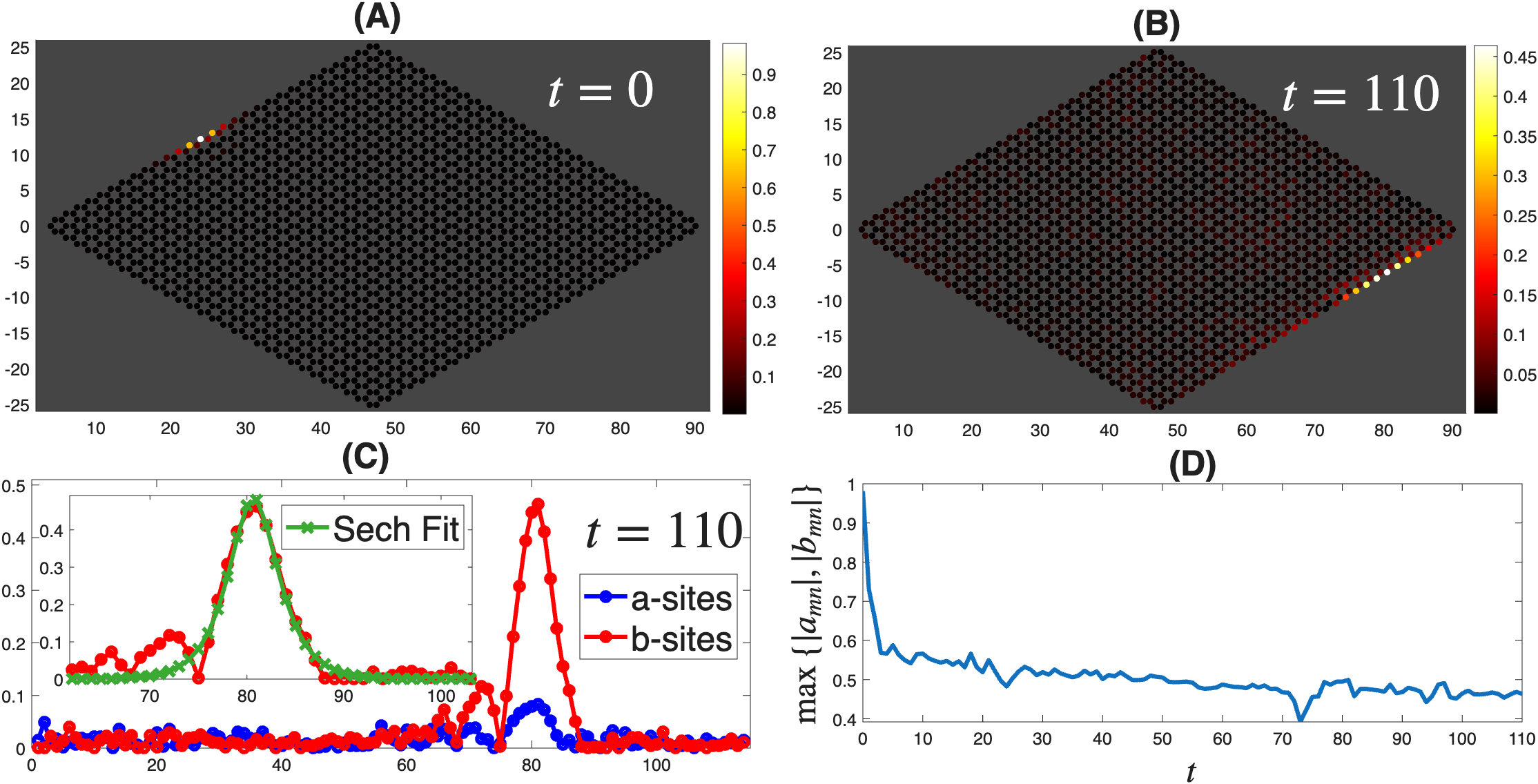}
\caption{(A)-(B) Evolution snapshots showing $|a_{mn}|,|b_{mn}|$ solutions of the ALH model (\ref{alhnonlin_ab}) using the  initial condition   in Fig.~\ref{alhNoMovie} with a sech envelope (\ref{sechIC}). (C) Side profile snapshot of the evolution highlighting a localized soliton-like solution. The inset shows a sech fit, $|b_\text{edge}| \approx 0.4856~\text{sech} %(0.438 m-35.33)
(0.438 m)$, of the main soliton-like hump. (D) Maximum magnitude of the solution as a function of time.   \label{alhSech}}
   \end{figure*}

Motivated by the previous results, we look for single-hump soliton-like modes. To do this, we modulate a linear edge mode by a secant hyperbolic profile;  this is the form predicted in the asymptotic results  described in \cite{Ablowitz2014}. That is, we take the same initial condition used in Figs.~\ref{nhNoMovie} and \ref{alhNoMovie},
and apply a secant hyperbolic envelope %of amplitude $A$ 
%%%%%%%%%%%%%%%%%%%%%%%%%%%%%%%%%%%%%%%%%%%%%%%%%%%%%%%%%%%%%%%%%%%%%
  \begin{equation}
       \label{sechIC}
       \begin{split}
       a_{mn}(0)=A\text{ sech}(m-m_0)a_ne^{2i\pi r} \\  b_{mn}(0)=A\text{ sech}(m-m_0)b_ne^{2i\pi r} ,
       \end{split}
   \end{equation}  
%%%%%%%%%%%%%%%%%%%%%%%%%%%%%%%%%%%%%%%%%%%%%%%%%%%%%%%%%%%%%%%%%%%%%
where $a_ne^{2i\pi r} , b_ne^{2i\pi r} $ correspond to the linear embedded edge mode, $A >0 $ is the amplitude of the  envelope, and $m_0$ is  a parameter that centers the envelope in the middle of the lattice edge.  
A summary of the findings with $A=1$, and $r=0.5$ is given in Fig.~\ref{alhSech}. After an initial transient period where the mode radiates energy and decreases amplitude, a steady solitary structure is observed to propagate around the edge. This soliton-like mode has nearly constant amplitude (see Fig.~\ref{alhSech}(D)) and a main profile that is well approximated by a secant hyperbolic profile. Similar to Fig.~\ref{alhNoTop}, the local Chern number is $C^L = +1$ in the lattice region. 

Upon closer inspection of the solitary profile, there does appear to be a small, but growing, dispersive tail. This indicates that a maximal balance, between dispersion and focusing nonlinearity, has not been achieved.
These results certainly suggest that traveling  solitons  may in fact exist in the ALH model, however we have not found any genuine solitons (constant form and velocity) yet. Nonetheless, these are promising results for the creation of chiral edge solitons and suggest this type of nonlinearity is more conducive to coherent nonlinear structures.

%%%%%%%%%%%%%%%

%%%%%%%%%%%%%%%%%%%%%%%%%%%%%%%%%%%

%%%%%%%%%%%%%%%%%%%%%%%%%%%%%%%%%%%%%%%%%%%%%%%%%%%%%%%%%%%%%%%
\subsection{Stationary Surface Solitons}
\label{stationary_soliton_sec}
%%%%%%%%%%%%%%%%%%%%%%%%%%%%%%%%%%%%%%%%%%%%%%%%%%%%%%%%%%%%%%%

In addition to the solutions described above, both the NH (\ref{nonlin_ab}) and ALH (\ref{alhnonlin_ab}) models support stationary (non-traveling) soliton solutions.  Since this manuscript is primarily focused on traveling modes, we leave the full discussion to Appendix~\ref{stationary_ALH_sec}. A spectral renormalization method, described in Appendix~\ref{specR}, is used to numerically compute solitons in the ALH model.

%In brief, a family of localized surface solitons (localized along an edge) exist, but their corresponding eigenvalues lies in the upper (lower) spectral gap regions for $\sigma = 1~(-1)$; \tij{double check this} \jtcc{JTC: Should we drop the negative case?}
 In brief, a family of localized surface solitons (localized along an edge) exist, but their corresponding eigenvalues lie in the upper spectral gap region; recall the band diagram given in Fig.~\ref{linearbands}(A). In general, these eigenmodes are extremely localized with the only nontrivial contributions coming from a  central lattice site and its 3 nearest neighbors (see Fig.~\ref{surfacemode}). These solutions are interesting in the context of the next section where we consider soliton interactions.

%%%%%%%%%%%%%%%%%%%%%%%%%%%%%%%%%%%%%%%%%%%%%%%%%%%%%%%%%%%%%%%   
\section{Interaction of Nonlinear Waves}
\label{interact_sec}
%%%%%%%%%%%%%%%%%%%%%%%%%%%%%%%%%%%%%%%%%%%%%%%%%%%%%%%%%%%%%%%

   In this final section we examine  interactions in the ALH model (\ref{alhnonlin_ab}) between the traveling edge modes in Sec.~\ref{alh_sect} and the stationary surface states in Sec.~\ref{stationary_soliton_sec}. In linear topological insulators, interactions are a superposition of solutions. As such, there is little opportunity for unexpected interaction dynamics to occur. Nonlinearity, on the other hand, affords us the chance to interact nonlinear coherent structures, such as solitons, and observe interesting dynamics. 
   
   Returning to the seminal  work of Zabusky and Kruskal \cite{ZabuskyKruskal65}, one of the first questions asked is whether solitons interact elastically with each other. Indeed, elastic interaction of nonlinear solitary states was part of
   the original definition of a soliton. Subsequently, these solitons were  found to be related to integrable  equations, which can be solved exactly via the inverse scattering transform \cite{Ablowitz1981}. On the other hand, inelastic collision of solitons have been observed in numerous works. These solutions %can 
   have many intriguing dynamics, such as different exiting velocities \cite{Kodama1987} and spiraling motion \cite{Rotschild2006}.

   To compare the different cases below, we fix the stationary power at $\mathbb{P}_{\text{ALH}} =2.7988 $ for the conserved power %energy 
   in (\ref{alh_mass}), corresponding to an eigenvalue of $\lambda=4.9$. Note that linear edge 
   eigenstates only exist for $\lambda \le 3$. The stationary modes shown correspond to stable modes discussed in Appendix~\ref{stationary_ALH_sec}. On the other hand, traveling waves are seeded by
   localized linear edge states, % edge solutions, 
   of the sort shown in Fig.~\ref{alhSech}, and will collide with the stationary edge mode.   
   Below we point out that varying the amplitude has an effect on the eventual dynamics,
   indicating the impact of 
   nonlinearity. This is different from the soliton collision mechanism discussed in \cite{Chong2016} which explores a phase difference between solitons to affect their interaction.  We have not yet been able to create  conditions necessary to interact two traveling waves. The reason is that  all chiral states  observed  so far  appear to travel at nearly the same speed and orientation, making interactions challenging.

\begin{figure*}
   \centering
\includegraphics[scale=.40]{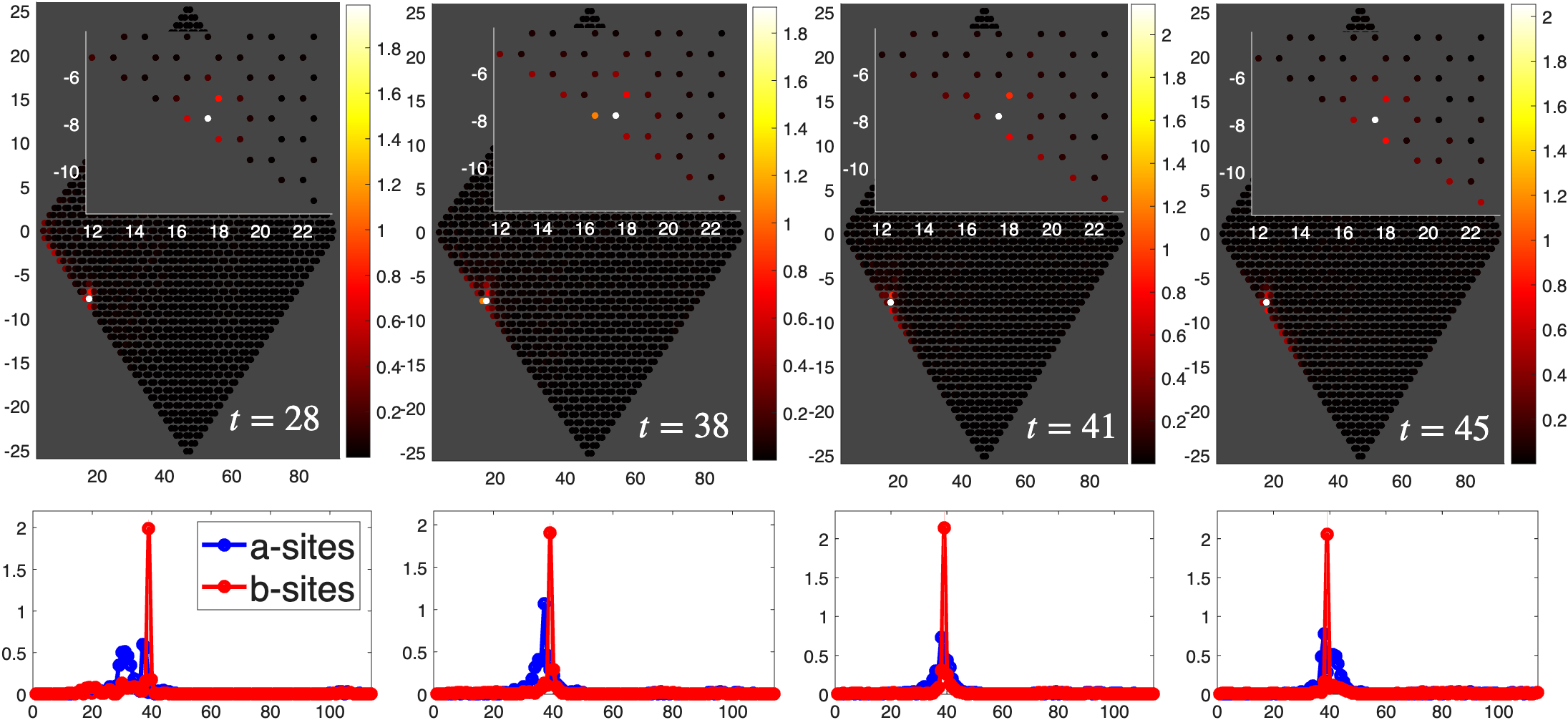}
\caption{Typical results of interacting a nonlinear traveling wave from Sec.~\ref{alh_sect} with a small amplitude ($A = 1$) secant hyperbolic envelope %of peak amplitude 1 
and a stationary mode from Appendix \ref{stationary_ALH_sec} with eigenvalue $\lambda=4.9$. (Top row) %shows the 
Snapshots of $|a_{mn}|,|b_{mn}|$ before, during, and after the interaction; there is little change to the traveling or stationary solitons. The inset shows a zoomed-in view. (Bottom row) Profile of the magnitude along the edge showing the collision. \label{alhS1}}
   \end{figure*}

   The first interaction is shown in Fig.~\ref{alhS1}. This corresponds to a traveling solitary wave with a relatively small amplitude, $A=1$ in Eq.~(\ref{sechIC}),  interacting with the stationary mode. The result looks nearly elastic: the incoming solitary wave passes through stationary soliton, with both keeping their initial amplitudes. We note that nearly elastic interactions have also been observed in nonlinear dislocated Floquet lattices \cite{Ivanov2020}. However, upon closer inspection, the interaction is not totally elastic. The stationary soliton has began a small cyclic motion between  the three nearest neighbor sites where energy oscillates around the central peak. This mircromotion resembles the bulk soliton states observed in nonlinear Floquet systems \cite{Mukherjee2020}.

\begin{figure*}
   \centering
\includegraphics[scale=.40]{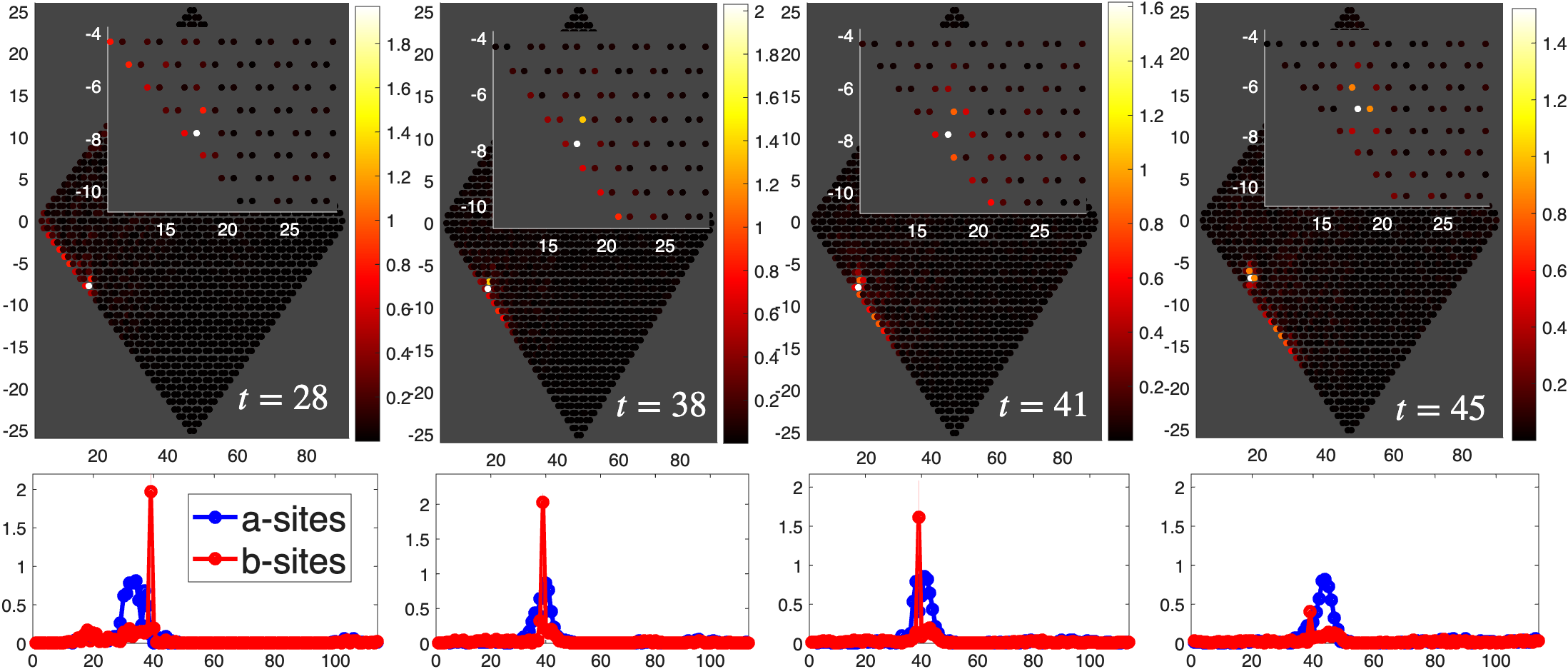}
\caption{Typical results of a nonlinear traveling wave %interacting a mode 
from Sec.~\ref{alh_sect} with a moderate amplitude ($A = 2$) secant hyperbolic envelope %of peak amplitude 2 
and a stationary mode from Appendix \ref{stationary_ALH_sec} with eigenvalue $\lambda=4.9$. (Top row) %shows the 
Snapshots of $|a_{mn}|,|b_{mn}|$ before, during, and after the interaction; the traveling mode moves the stationary soliton away from the edge, closer to the bulk. The inset shows a zoomed-in view. (Bottom row) %shows the 
 Profile of the magnitude along the edge showing the collision; the amplitude of the stationary soliton ($b-$sites) appears to decrease since it is no longer located on the edge.  \label{alhS2}}
   \end{figure*}

    Next, we consider a moderately large amplitude wave, $A=2$ in Eq.~(\ref{sechIC}). A few snapshots of this interaction are shown in Fig.~\ref{alhS2}. This mode has a more significant impact on the stationary mode. As a result of the interaction, the stationary soliton is pushed off the surface edge and closer to the bulk, by one site. The post-interaction stationary soliton also exhibits a local cyclic motion.

\begin{figure*}
   \centering
\includegraphics[scale=.40]{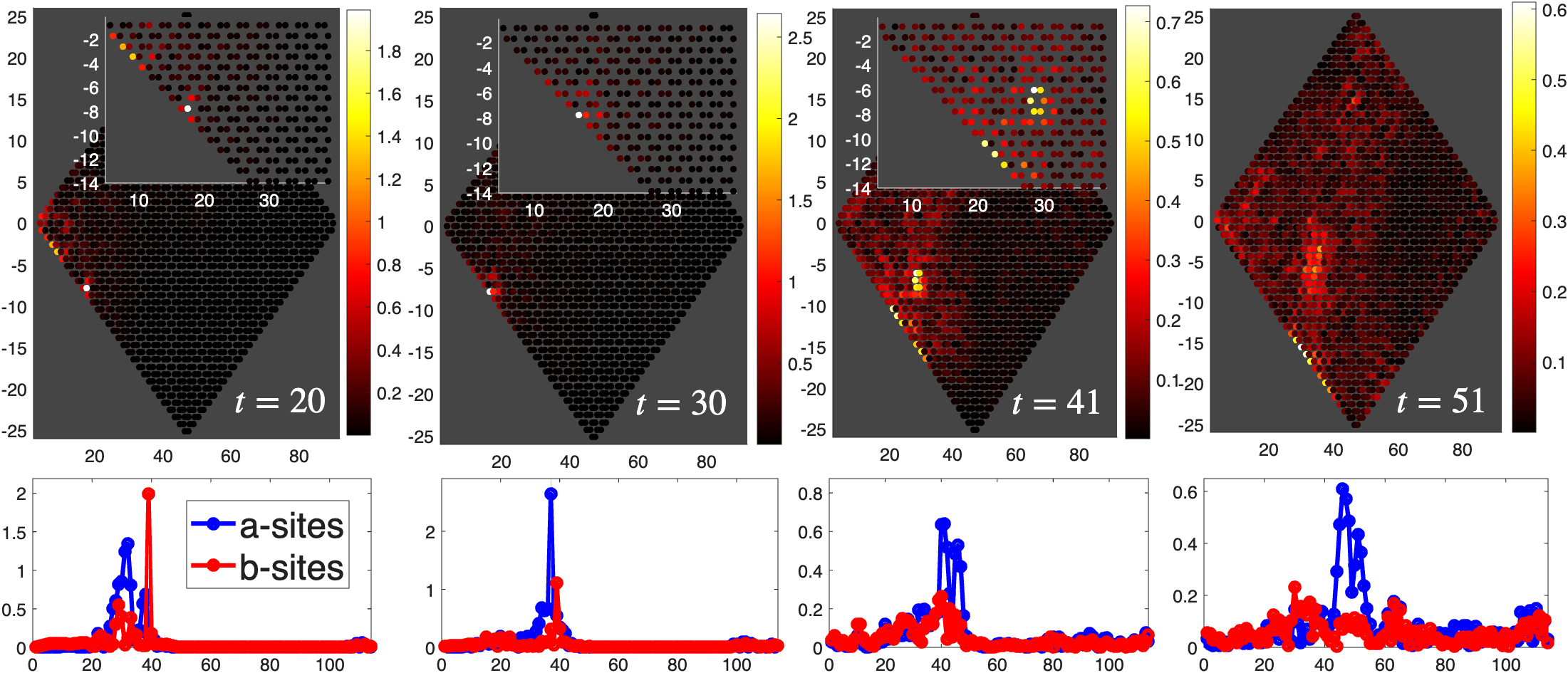}
\caption{Typical results of a nonlinear traveling wave %interacting a mode 
from Sec.~\ref{alh_sect} with a large amplitude $($A = 4$)$ secant hyperbolic envelope %of peak amplitude 4 
and a stationary mode from Appendix \ref{stationary_ALH_sec} with eigenvalue $\lambda=4.9$. (Top row) %shows the 
Snapshots of $|a_{mn}|,|b_{mn}|$ before, during, and after the interaction; the traveling mode effectively disintegrates the stationary soliton and yields radiation. The inset shows a zoomed-in view. (Bottom row) %shows the 
 Profile of the magnitude along the edge showing the collision and subsequent dissolution of the stationary soliton. \label{alhS4}}
   \end{figure*}

    A final set of interaction simulations is
    shown in Fig.~\ref{alhS4}, corresponding to the largest amplitude traveling wave we consider, $A=4$ in Eq.~(\ref{sechIC}).  The interaction is clearly inelastic and fatal for the stationary soliton. The high power traveling wave disintegrates the stationary soliton, completely radiating it away into the bulk. 

     It should be pointed out that all of these interactions correspond to an inter-gap interaction. That is, the traveling modes corresponding to mid-gap frequencies (see Fig.~\ref{linearbands}(A)) interacting with  stationary solitons whose eigenvalues lie in the upper band gap (see Fig.~\ref{surfacemode}). We conjecture this is why the final case does not result in a merging of the two states: they are not supported at the same frequencies.
    
    The local topology of this section appears to give little insight into the eventual dynamics. The regions of local topology, as measured by the spectral localizer (\ref{lLam}), resemble those of the linear and ALH models in Secs.~\ref{lChern} and \ref{alh_sect}, respectively. Namely, for spatial regions inside the lattice and energy eigenvalues inside the linear band gaps, the local Chern number is $C_L = +1$. Outside these regions, the local Chern number tends to be null. These results motivate the need for further study of nonlinear waves and their relationship to topological protection.  %interaction.  

    We conjecture a possible application for this sort of dynamics. The strongly nonlinear waves depicted in Fig.~\ref{alhS4} function as a ``cleaner'' of surface stationary modes. That is, they essentially remove localized filaments from the interface. %or clean surface states. 
    This mechanism could be useful in cleaning the surface of photonic crystals, especially in places that are difficult to access directly, or are more accessible by traversing 
    a topological interface. 
    
    As a final comment, we experimented with a few other models that have similar nonlinearity 
 and found similar results. For example, only modulating the intensity on nearest neighbor terms had a similar outcome: conductive chiral states. %These observations motivate future  systems that can apply nonlinearity in a (weakly) nonlocal fashion.

%%%%%%%%%%%%%%%%%%%%%%%%%%%%%%%%%%%%%%%%%%%%%%
%%%%%%%%%%%%%%%%%%%%%%%%%%%%%%%%%%%%%%%%%%%%%%

  % \begin{figure}
   %\centering
%\includegraphics[scale=.20]{ALH_INT_AMP_8_19_25.png}
%\caption{The top figure shows the max amplitude of the modes in Fig.  (\ref{alhS4}). The bottom figure shows absolute value of the difference between the initial power(energy) in blue(red) and the power(energy) at time $t$.   \label{alhInt}}
%\end{figure}

\section{Conclusions}
\label{conclude_sec}
%%%%%%%%%%%%%%%%%%%%%%%%%%%%%%%%%%%%%%%%%%%%%%%%%%%%%%%%%%%%%%%

A nonlinear tight-binding model with two conserved quantities that %and 
resembles the well-known Haldane model modulated by an intensity-dependent term $1 + \sigma |\psi_{mn}|^2$ was introduced. The eventual form of the nonlinearity %looks similar to 
resembles 
that of the integrable Ablowitz-Ladik lattice equation, hence we refer to this as the Ablowitz-Ladik-Haldane model. This model is found to better support edge currents  than on-site, Kerr-type nonlinearity, especially at large power. The solutions exhibit inelastic interactions and so  it is not believed that these models are integrable.

This work serves as motivation for future direction in nonlinear topological insulators. The ALH model indicates that lightly nonlocal coupling is conducive to supporting chiral edge states. Several works have give physical motivations for this sort of nonlinearity (see \cite{Konotop1997,Aceves1996,Haldane1982,Ishimori1982,Castro2025}). This indicates that it is possible to realize Ablowitz-Ladik type nonlinearity and observe the solitonlike states shown in this work. %, provided there is an additional effective magnetic field to break time-reversal symmetry \cite{Ablowitz2024}.}

Applications such as a %self 
soliton-cleaning mechanism at high power and solitonic optical communication \cite{Hasegawa1995} appear to be viable in these setups.

\section*{Acknowledgments}
We acknowledge helpful discussions with Alexander Cerjan and Stephan Wong. This work was supported by AFOSR research Grant No. FA9550-23-1-0105.

\appendix

%%%%%%%%%%%%%%%%%%%%%%%%%%%%%%%%%%%%%%%%%%%%%%%%%
\section{Varying the Nonlinearity of the Nonlinear Haldane Model}
\label{weak_sect}
%%%%%%%%%%%%%%%%%%%%%%%%%%%%%%%%%%%%%%%%%%%%%%%%%

In this section we detail the results of decreasing the strength of the nonlinearity in the NH model (\ref{nonlin_ab}).  We  use the same standard parameters as in Sec. \ref{Travel_Sec}:  $M=0$,  $t_1=1$, $t_2=0.1$ and $\phi=\frac{\pi}{2}$, except now we set $\sigma=0.1$.

\begin{figure*}
\centering
\includegraphics[scale=.40]{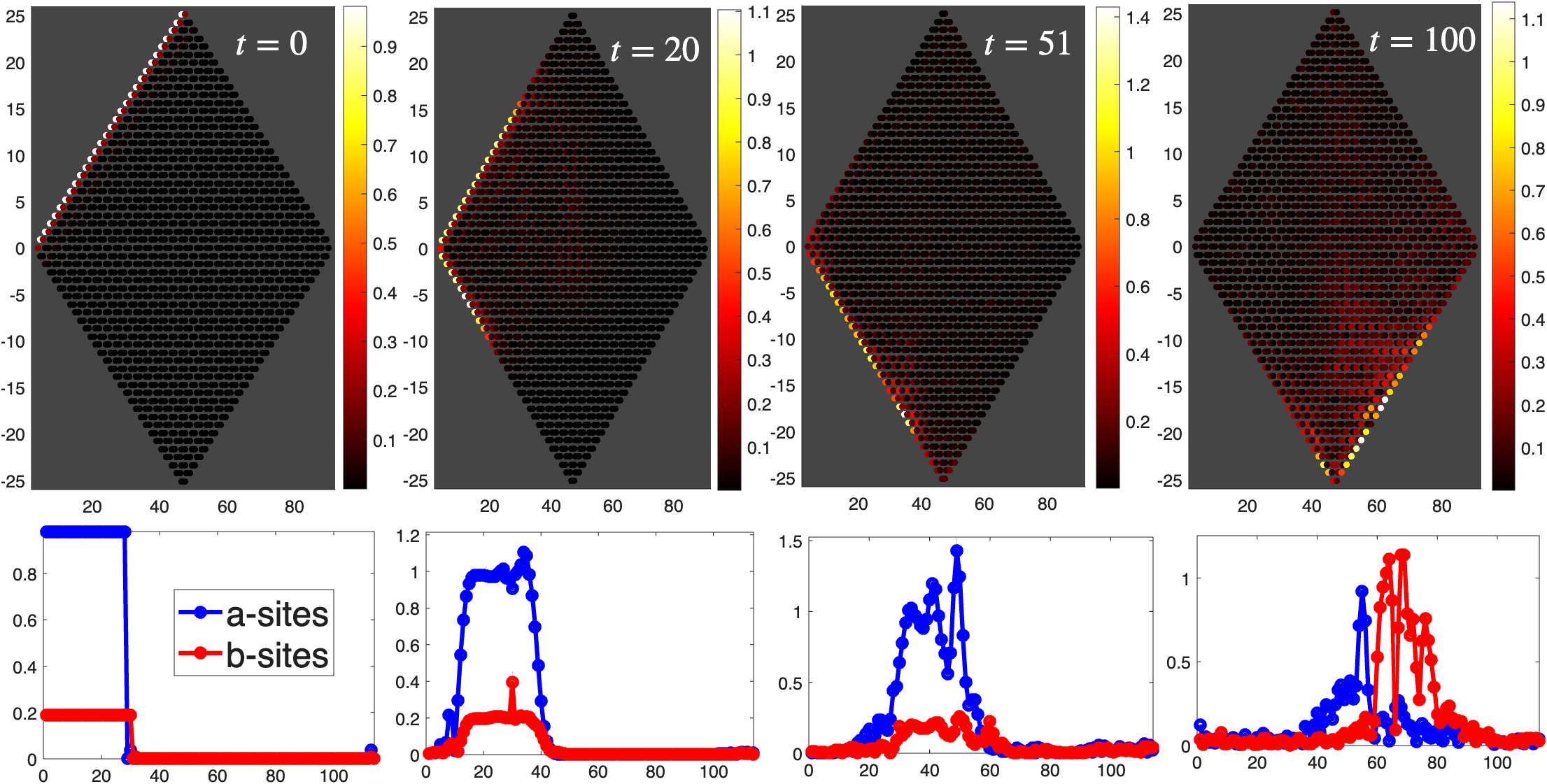}
\caption{(Top row) Evolution of $|a_{mn}|,|b_{mn}|$ in the weakly nonlinear ($\sigma = 0.1$) NH model (\ref{nonlin_ab}) with %\jtc{the} 
a linear ($\lambda = 0$) edge mode initially embedded at $n =0$. The edge mode  radiates considerably less relative to the strong nonlinearity case ($\sigma = 1$) in Fig.~\ref{nhNoMovie}.
 (Bottom row) Edge %Side 
 profile snapshots of the evolution shown in the top row. These magnitudes correspond to the outermost lattice sites, along the perimeter of the domain.  \label{nhWeakmovie}}
\end{figure*}

The evolution of a weakly nonlinear mode in the NH model is shown in Fig.~\ref{nhWeakmovie}. Compared to Fig.~\ref{nhNoMovie} (where $\sigma = 1$), significantly less radiation is observed. The peak amplitude, is not observed to significantly decrease over the same time period. Once again, power (\ref{nhmass}) and energy (\ref{nh_ham}) are conserved. 
%there is no loss of power or energy.

These results are to be expected as $\sigma \to 0$ in the %NH Haldane model, \jtcc{since it} approaches the Haldane model \jtcc{in this limit}.
NH model, since it approaches the (linear) Haldane model in this limit.
If a slowly-varying envelope is attached to the linear edge envelope, we will reproduce results similar to \cite{Ablowitz24} and enter the asymptotic regime described in \cite{Ablowitz2014,Ablowitz2021}. Namely, when  the spatial dispersion of edge
  profiles balances \textit{weak} self-focusing nonlinearity, to leading-order, edge solitons exist.

%%%%%%%%%%%%%%%%%%%%%%%%%%%%%%%%%%%%%%%%%%%%%%%%%%%%%%%%%%%%%%%%%%%%%%
%%%%%%%%%%%%%%%%%%%%%%%%%%%%%%%%%%%%%%%%%%%%%%%%%%%%%%%%%%%%%%%%%%%%%%

\section{Stationary modes of the Ablowitz-Ladik-Haldane equation}
\label{stationary_ALH_sec}

 In this section we discuss finding families of surface modes for the ALH  model (\ref{alhnonlin_ab})  using   spectral renormalization method described in Appendix~\ref{specR}.

\begin{figure*}
   \centering
\includegraphics[scale=.45]{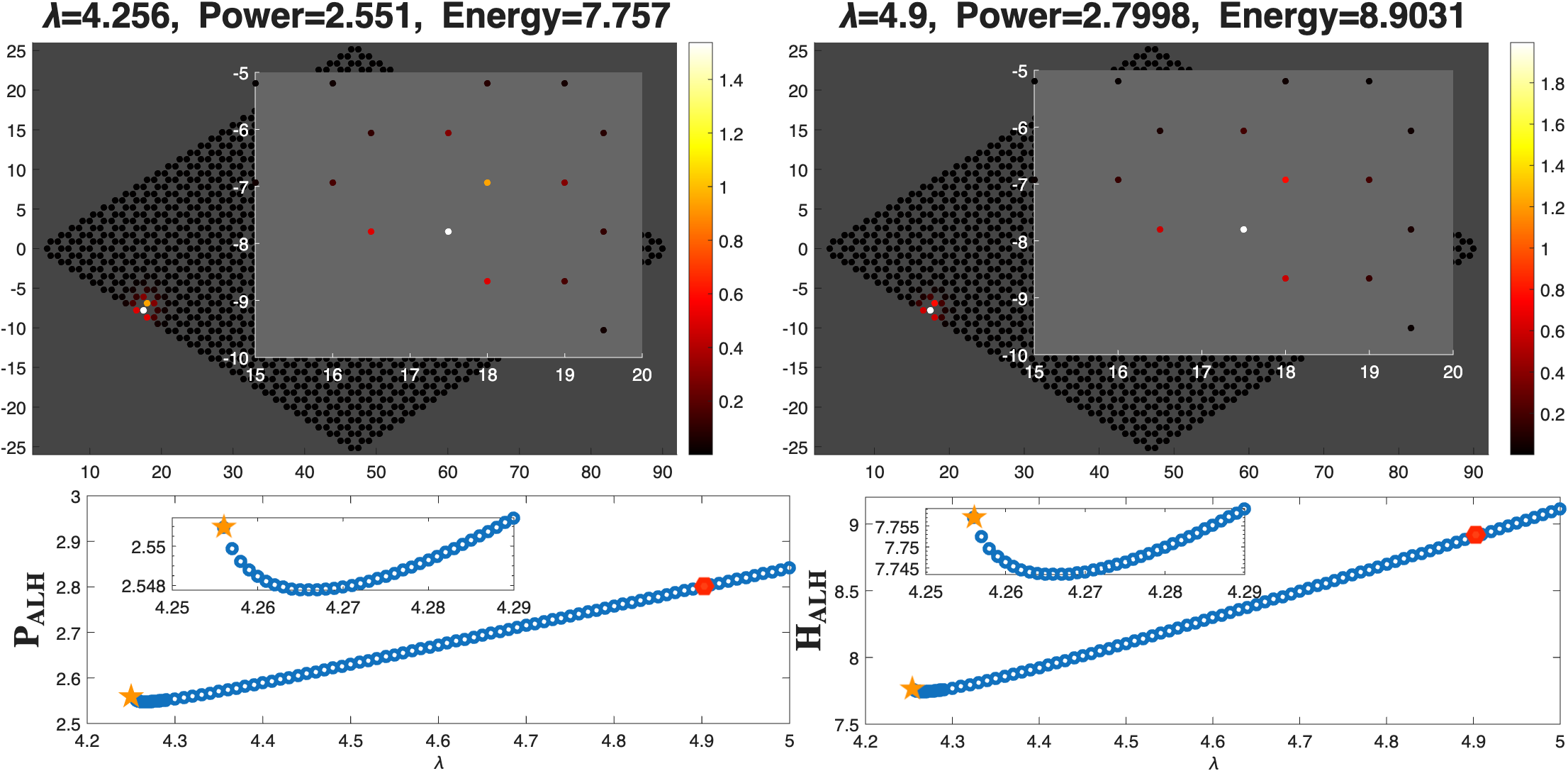}
\caption{(Top row) Typical surface solitons  obtained by the spectral renormalization algorithm (Appendix~\ref{specR})  for parameters $M = 0$, $t_1=1$, $t_2=0.1$ and $\phi=\pi/2$ on a $30 \times 30$ lattice. The inset highlights the localized structure of the nonlinear edge mode. (Bottom row) Power (\ref{alh_mass}) and energy (\ref{alh_ham}) curves as a function of  the soliton  eigenvalue. The inset highlights a  turning point (minimum) in the curves near the linear band edge.  The orange star corresponds to the mode on the top left and the red hexagon corresponds to the mode on the top right. \label{surfacemode}}
   \end{figure*}

  A few typical surface modes are shown in Fig.~\ref{surfacemode} (top row). Both are exponentially localized near the edge with the largest peak in the next-to-last zig-zag row. That is, the largest peak is not in the site closest to the boundary, unlike the standard linear zig-zag edge states \cite{Ablowitz2013}. The power (\ref{alh_mass}) and energy (\ref{alh_ham}) curves as a function of eigenvalue for this family of surface solitons are shown in Fig.~\ref{surfacemode} (bottom row).  We note that both the power and energy curves possess  a  minimum (see insets in Fig.~\ref{surfacemode} (bottom row)). Anecdotally, we observe a VK-instability criteria \cite{VK1973} where soliton with power slope $d \mathbb{P}/d \lambda >0~ (<0)$ corresponds to stable (unstable) solitons. Direct numerical simulations confirm this intuition. A comprehensive study of these soliton modes is outside the scope of this work. We observe a zero local Chern number for all energies in this upper gap region. As such, we do not anticipate traveling edge soliton behavior.

Lastly, we note the existence of several other families of stationary solitons in the upper gap. These include corner modes (localized in a corner of the lattice) and bulk solitons (localized in the interior, far from the edges). However, we do not consider them here.
%%%%%%%%%%%%%%%%%%%%%%%%%%%%%%%%%%%%%%%%%%%%%%%%%%%%
%%%%%%%%%%%%%%%%%%%%%%%%%%%%%%%%%%%%%%%%%%%%%%%%%%%%%
%%%%%%%%%%%%%%%%

%%%%%%%%%%%%

%%%%%%%%%%%%%%%%%%%%%%%%%%%%%%%%%%%%%%%%%%%%%%%%%%%%%%%%
\section{Spectral Renormalization}
\label{specR}
%%%%%%%%%%%%%%%%%%%%%%%%%%%%%%%%%%%%%%%%%%%%%%%%%%%%%%%%

In this section details on  a spectral  renormalization method \cite{Abowitz2005} used to compute the family of stationary gap soliton modes for the ALH equation (\ref{alhnonlin_ab})  in Sec.~\ref{stationary_soliton_sec} and Appendix~\ref{stationary_ALH_sec} are given. % The algorithm for both the NH (\ref{nonlin_ab}) and ALH (\ref{alhnonlin_ab}) equations is similar.

To begin, consider %Look for 
time-harmonic, one-dimensional solutions of the form ${\bf c}(t) =  \boldsymbol{\psi} e^{ - i \lambda t}$, where $ \boldsymbol{\psi} = ({\bf a_{mn}} | {\bf b_{mn}})$ is the time-independent edge eigenstate. The eigenmodes at $a_{mn}$ and $b_{mn}$ for $m,n = 1,\dots, N$ and are stored in ${\bf a_{mn}}$  and ${\bf b_{mn}}$, respectively. This gives nonlinear eigenvalue problems of the form 
%%%%%%%%%%%%%%%%%%%%%%%%%%%%%%%%%%%%%%%%
\begin{equation}
\label{nlEigProb}
(H_\text{lin} + N[| \boldsymbol{\psi}|^2]) \boldsymbol{\psi} = \lambda \boldsymbol{\psi} ,
\end{equation}
%%%%%%%%%%%%%%%%%%%%%%%%%%%%%%%%%%%%%%%%
where $H_\text{lin}$ is the linear Haldane Hamiltonian matrix and $N[|\boldsymbol{\psi}|^2]$ contains %is 
the nonlinear terms. We note that for  the on-site model in (\ref{nonlin_ab}), $N[|\boldsymbol{\psi}|^2] =  \sigma \text{diag}(|\boldsymbol{\psi}|^2)$ and in the AL model in (\ref{alhnonlin_ab}), $N[|\boldsymbol{\psi}|^2] =\sigma \text{diag}(|\boldsymbol{\psi}|^2)H_\text{lin}$.

The essence of the spectral renormalization method is to look for a solution of the form $\boldsymbol{\psi} =\gamma \boldsymbol{\phi}$ for some renormalization factor $\gamma >0 $. Using this in (\ref{nlEigProb}) gives 
% %%%%%%%%%%%%%%%%%%%%%%%%%%%%%%%%%%%%%%%%%%%%
% \begin{equation}
%        \label{renorm_eig1}
%        (L+N_m(\gamma d))\gamma d=\lambda \gamma d
%    \end{equation}
% %%%%%%%%%%%%%%%%%%%%%%%%%%%%%%%%%%%%%%%%%%%%
%which yields 
%%%%%%%%%%%%%%%%%%%%%%%%%%%%%%%%%%%%%%%%%%%%
 \begin{equation}
       \label{renorm_eig2}
       H_\text{lin} \boldsymbol{\phi} +\gamma^2N[|\boldsymbol{\phi}|^2] \boldsymbol{\phi} =\lambda  \boldsymbol{\phi} ,
   \end{equation}   
%%%%%%%%%%%%%%%%%%%%%%%%%%%%%%%%%%%%%%%%%%%%
   which we can rewrite as 
%%%%%%%%%%%%%%%%%%%%%%%%%%%%%%%%%%%%%%%%%%%%t
\begin{equation}
       \label{renorm_eig3}
       (H_\text{lin}-\lambda I ) \boldsymbol{\phi} =-\gamma^2 N[|\boldsymbol{\phi}|^2] \boldsymbol{\phi} . 
   \end{equation}
%%%%%%%%%%%%%%%%%%%%%%%%%%%%%%%%%%%%%%%%%%%%
   If $\lambda$ is not in the spectrum of $H_\text{lin}$, we have that $(H_\text{lin}-\lambda I)$ is invertible and thus 
   \begin{equation}
       \label{renorm_eig4}
       \boldsymbol{\phi} =- \gamma^2 (H_\text{lin}-\lambda I )^{-1} N[|\boldsymbol{\phi}|^2] \boldsymbol{\phi} 
   \end{equation}
%%%%%%%%%%%%%%%%%%%%%%%%%%%%%%%%%%%%%%%%%%%%
is well-defined.

Equation (\ref{renorm_eig4}) can be viewed as a fixed-point method of the form $\boldsymbol{\phi} = F(\boldsymbol{\phi} )$. This then suggests the iteration
%%%%%%%%%%%%%%%%%%%%%%%%%%%%%%%%%%%%%%%%
   \begin{equation}
       \label{renorm_eig_it}
       \boldsymbol{\phi}^{(k+1)} =- (\gamma^{(k)})^2 (H_\text{lin}-\lambda)^{-1} N[|\boldsymbol{\phi}^{(k)}|^2] \boldsymbol{\phi}^{(k)} ,
   \end{equation}
%%%%%%%%%%%%%%%%%%%%%%%%%%%%%%%%%%%%%%%%
where $k$ denotes the iteration index. Without the renormalization factor, this method will typically converge to zero or blow up. 

For a given gap eigenvalue $\lambda$,  the renormalization factor, is determined by multiplying (\ref{renorm_eig3}) by $\boldsymbol{\phi}^\dagger$ and solving for $\gamma^2$  we get 
%%%%%%%%%%%%%%%%%%%%%%%%%%%%%%%%%%%%%%%%
   % \begin{equation}
   %     \label{lam_eq}
   %     \lambda=\frac{d^\dagger Ld+|\gamma|^2d^\dagger N_m(d)d}{d^\dagger d}
   % \end{equation}
%%%%%%%%%%%%%%%%%%%%%%%%%%%%%%%%%%%%%%%%
%and by isolating $|\gamma|^2$ we have 
%\begin{equation}
 %      \label{gam_eq}
  %     |\gamma|^2=\frac{\lambda d^\dagger d-d^\dagger Ld}{d^\dagger N(d)d}
   %\end{equation}
% Then since, Power$=||c||^2=|\gamma|^2||d||^2=$constant, we have that

% \begin{equation}
%        \label{pow_eq}
%        |\gamma|^2=\frac{\text{Power}}{||d||^2}
%    \end{equation}
%    Then given suitable initial guesses $d_0$ and Power, we can utilize, (\ref{pow_eq}) to form recurrence relations from (\ref{lam_eq}) and(\ref{renorm_eig4}), 
% \begin{equation}
%        \label{pow_rr}
%        |\gamma_n|^2=\frac{\text{Power}}{||d_n||^2}
%    \end{equation}
% \begin{equation}
%        \label{lam_rr}
%        \lambda_n=\frac{d_n^\dagger Ld_n+|\gamma_n|^2d_n^\dagger N_m(d_n)d_n}{d_n^\dagger d_n}
%    \end{equation}
%    \begin{equation}
%        \label{rec_rel}
%        d_{n+1}=-(L-\lambda_nI)^{-1}|\gamma_n|^2N_m(d_n)d_n
%    \end{equation}
%    Equations (\ref{pow_rr})-(\ref{rec_rel}) form the framework of an algorithm for converging to solutions of (\ref{eigProb}).
   % Alternatively, we can solve (\ref{renorm_eig4}) for $|\gamma|^2$ and by multiplying (\ref{renorm_eig4}) by $d^\dagger$ and isolating $|\gamma|^2$ we get
%%%%%%%%%%%%%%%%%%%%%%%%%%%%%%%%%%%%%%%%%%
   \begin{equation}
       \label{gam_eq}
       \gamma^2=\frac{\lambda \boldsymbol{\phi}^\dagger \boldsymbol{\phi}-\boldsymbol{\phi}^\dagger H_\text{lin}\boldsymbol{\phi}}{\boldsymbol{\phi}^\dagger N[|\boldsymbol{\phi}|^2]\boldsymbol{\phi}} ,
   \end{equation}
%%%%%%%%%%%%%%%%%%%%%%%%%%%%%%%%%%%%%%%%%%
which is well-defined as long as $\lambda > \rho(H_\text{lin})$, where $\rho(H_\text{lin})$ is the spectral radius of $H_\text{lin}$, %since the denominator is positive provided $\boldsymbol{\phi} \not= {\bf 0}$. \jtcc{JTC: How do we know the denominator is positive for ALH?} 
and the denominator is positive provided $\boldsymbol{\phi} \not= {\bf 0}$, which it is for suitable initial guesses. Our code checks that %for the method has a check that ensures 
the denominator is positive for each iteration. In terms of the iterative method in Eq.~(\ref{renorm_eig_it}), $\gamma^2$ at iteration $k$ is 
%%%%%%%%%%%%%%%%%%%%%%%%%%%%%%%%%%%%%%%%%%
   \begin{equation}
       \label{gam_eq2}
       (\gamma^{(k)})^2=\frac{\lambda( \boldsymbol{\phi}^{(k)} )^\dagger \boldsymbol{\phi}^{(k)}-(\boldsymbol{\phi}^{(k)})^\dagger H_\text{lin}\boldsymbol{\phi}^{(k)}}{(\boldsymbol{\phi}^{(k)})^\dagger N[|\boldsymbol{\phi}^{(k)}|^2]\boldsymbol{\phi}^{(k)}} .
   \end{equation}
%%%%%%%%%%%%%%%%%%%%%%%%%%%%%%%%%%%%%%%%%%

%     Then, given suitable initial guesses 
     %\jtc{
     %%%%%%%%%%%%%%%%%%%%%%%%%%%%%%%%%%%%%%%%%%%
    % \begin{equation}
     %    \label{surface_initial_guess}
            %  \boldsymbol{\phi}^{(0)} =({\bf e_j}|0.1{\bf e_j})
    % \end{equation}
     %%%%%%%%%%%%%%%%%%%%%%%%%%%%%%%%%%%%%%%%%%%
     %}
     %\tij{where ${\bf e_j}$ is a basis vector}, and $\lambda$ in the upper band gap, we can utilize (\ref{renorm_eig_it}) and (\ref{gam_eq2}) to iterate until convergence. A typical initial guess  consists of a localized function located near the surface. The method iterates unill $||\boldsymbol{\phi}^{(k+1)}-\boldsymbol{\phi}^{(k)}||$ is acceptably small. 

     %Then, given suitable initial guesses, and $\lambda$ in the upper band gap, we %can 
     %utilize (\ref{renorm_eig_it}) and (\ref{gam_eq2}) sequentially to iterate until convergence is reached. A typical initial guess consists of picking an $a$-site on the surface and setting it equal to 1, while a neighboring %its corresponding 
     %$b$-site is set  to 0.1, and all other $a$ and $b$ sites are set to 0. This process also works by setting surface $b$-site equal to 1 and nearby %corresponding 
     %$a$-site equal to 0.1; there are no other nonzero sites in the initial guess. The method iterates until $||\boldsymbol{\phi}^{(k+1)}-\boldsymbol{\phi}^{(k)}||$ is acceptably small. 
     
      To generate this family of solutions we take a simple localized initial guess with %of the form %given above and  set 
 $b_{1,10}=1$ and $a_{1,10}=0.1$ and all other values fixed to zero. The method iterates until $||\boldsymbol{\phi}^{(k+1)}-\boldsymbol{\phi}^{(k)}||$ is acceptably small.  The modes we converge to correspond to eigenvalues $\lambda$ deep in the upper band  gap (well away from the linear band edge). After converging to a solution, we then decrease $\lambda$ (move closer to the linear band edge) and solve again except now as an initial guess we use the perviously obtained surface mode. We repeat the process iteratively until the spectral renormalization iteration %process 
 fails to converge to a surface mode. When the method fails, we find that it often %the solution 
 jumps to another family of solutions, e.g. %usually some 
 a hybrid edge-bulk soliton.

\end{document}